\documentstyle[aps,prl,multicol,epsfig]{revtex} 
\begin{document}
\title{ Zero temperature transition from d-wave superconductor to 
        underdoped regime }
\author{ Jinwu Ye $^{1,2}$ and A. Millis $^{1,3}$  }
\address{ $^1$Institute for Theoretical Physics, University of California,
    Santa Barbara, CA, 93106 \\
   $^2$ Department of Physics, University of Houston, TX, 77204 \\
   $^3$ Department of Physics, Rutgers university, Piscataway,
    New Jersey, 07974 }
\date{\today}
\maketitle
\begin{abstract}
   By using mutual flux-attaching singular gauge transformations,
  we derive an effective action describing the zero temperature
  quantum phase transition from d-wave superconductor to underdoped regime.
  In this effective action, quantum fluctuation generated vortices
  couple to quasi-particles
  by a mutual statistical interaction with statistical angle $ \theta $,
  the vortices are also interacting with each other by
  long range interactions due to charge fluctuation.
  Neglecting the charge fluctuation, we find a fixed line characterized
  by $ \theta $ and calculate the universal
  spinon, vortex and mutual Hall drag conductivities
  which continuously depend on $ \theta $. Implications for double layer
  quantum Hall systems are given. When incorporating the charge
  fluctuation, we emphasize the importance of keeping the periodicity of
  the mutual CS term at $ \theta= \pm 1/2 $ to get the correct critical
  behaviors and point out possible future directions.
  The connection with $ Z_{2} $ gauge theory is discussed.

\end{abstract}
\begin{multicols}{2}
\section{ Introduction}

   In this paper, we are trying to study the nature of zero 
  temperature quantum phase transition from d-wave superconductor
  at $ x > x_{c} $ to the underdoped regime at $ x < x_{c} $ of
  the high temperature superconductors ( Fig.1).

\vspace{0.25cm}

\epsfig{file=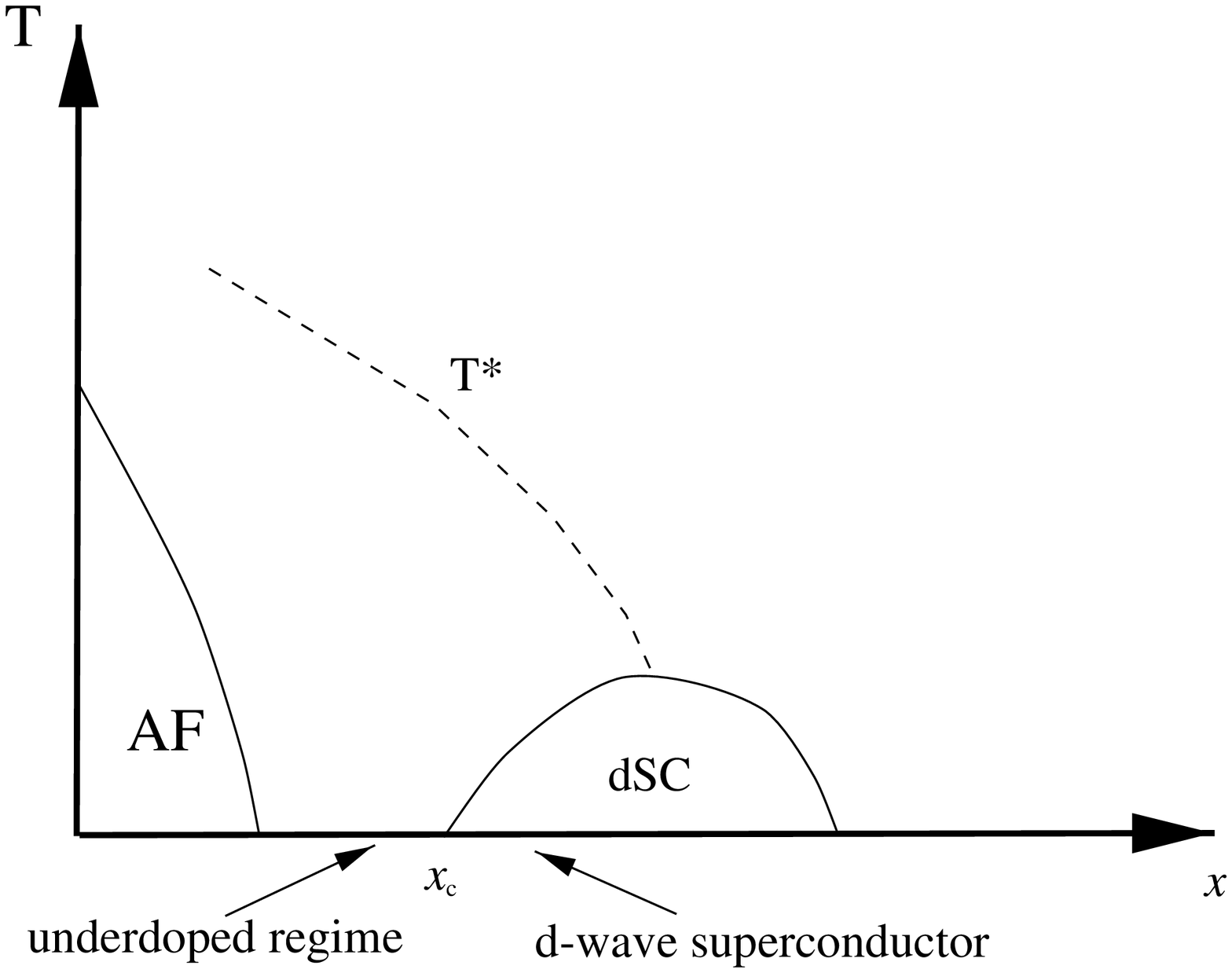,width=3.2in,height=2in,angle=0}

{\footnotesize {\bf Fig 1:} The temperature ( T ) versus doping ( $ x $ )
  diagram of high $ T_{c} $ cuprates.}

\vspace{0.25cm}

    It is well known that inside the superconductor phase $ x>x_{c} $,
  there are low energy
  quasi-particle excitations near the four nodes of the Fermi surface,
  the quantum phase fluctuation of the Cooper pair condensates are suppressed,
  the positive and negative vortices are bound together.
  As the doping decreases from the right to $ x_{c} $, the phase fluctuation
  increases. At the underdoped regime
  $ x < x_{c} $, the quantum phase fluctuations are so strong that
  they generate free $ hc/2e $ vortices, destroy the long range
  phase coherence of the d-wave superconductor, 
  therefore the superconducting ground state.
  However, the local short-range pairing still exists, the low energy
  quasi-particles at the four nodes remain.
  This quantum phase transition is driven by the condensation of
  $ hc/2e $ vortices. Near the transition, there are strong coupling
  between the quasi-particles and the phase windings of $ hc/2e $ vortices. 

  A lot of work has been done on a closely related problem where
 quasi-particles are coupled to the vortices generated by external
 magnetic field inside the superconducting state 
 \cite{volovik,hirs,and,sing,bert1,static}. 
 In the following, we will first review these work, then will extend
 the methods developed in these work to study this quantum phase transition
 where the quasi-particles are coupled to the quantum fluctuation
 generated vortices.

 Employing semi-classical approximation, Volovik pointed out that  
 the circulating supercurrents around vortices
 induce Doppler energy shift to the quasi-particle spectrum, which leads to
 a finite density of states at the nodes \cite{volovik}. 
 This effect ( Volovik effect) has been employed to
 attempt to explain the experimental observations
  \cite{ong} of longitudinal thermal
 conductivity $ \kappa_{xx} $ in \cite{hirs}.

 Starting from BCS Hamiltonian, Anderson \cite{and} employed the
 first single-valued singular
 gauge transformation to study quasi-particle dynamics in the mixed state.
 Unfortunately, Anderson
 made an incorrect mean field approximation which violates the "Time-Reversal"
 symmetry \cite{static}, therefore leads to the incorrect conclusion that
 there is Landau level quantization of energy levels of the quasi-particle. 
 Franz and Tesanovic (FT) employed a  different  single-valued 
 singular gauge transformation
 to map the quasi-particle in a square vortex lattice state to Dirac fermion
 moving in an effective periodic scalar and vector potential with {\em zero}
 average and studied the quasi-particle spectrum numerically \cite{sing,bert1}.
 They did not see the signature of Landau level quantization in their numerical
 calculations

 Unfortunately, in the vortex lattice state, the vector potential only
 provides a periodic potential instead of scattering quasi-particles,
 its effect on the energy spectrum and other physical quantities
 is not evident. This motivated Ye to study the quasi-particle transport
 in the disordered vortex state where the random AB
  phase scattering may show its important effects.
  In Ref.\cite{static}, Ye observed that because infinite thin $ hc/2e  $
  vortices do {\em not} break the T reversal symmetry, any correct
  mean field theory should respect this T symmetry.
  He showed exactly there is no Landau level quantization due to this
  T symmetry.
  He applied the singular transformation in the FT gauge to {\em disordered}
  vortex state and found that
  the long-range logarithmic interaction between vortices suppresses the
  fluctuation of superfluid velocity (scalar potential), but does {\em not}
   affect the fluctuation of the internal gauge field.
   Therefore the scalar field acquires a " mass " determined by the vortex
   density, but the gauge field remains  " massless ".
  The quasi-particle scattering from the "massless" internal
  gauge field dominates over those from the well-known "massive"
  Volovik effect and the non-magnetic scattering
  at sufficient high magnetic field. This dominant scattering mechanism
   is a purely quantum mechanical effects which
   was completely missed by all the previous
   semi-classical treatments \cite{hirs}.
  In fact. it is responsible for the
  behaviors of both $ \kappa_{xx} $ and $ \kappa_{xy} $ in high magnetic field
  observed in the experiments \cite{ong}.  

     When the vortices are generated by quantum fluctuations or
 thermal fluctuations, themselves are moving around, new physics may
 arise. From Newton's
 third law, intuitively, the vortices must also feel the counter-acting AB
 phase coming from the quasi-particles. Therefore there is a mutual AB phase
 scattering between vortices and quasi-particles.
  In this paper, we enlarge the {\em one} singular gauge transformation
 developed in \cite{and,sing,static} to {\em two} mutual 
 flux-attaching singular 
 gauge transformations. We then apply them to study quasi-particles coupled to
 moving vortices near the zero temperature quantum critical point.
  In canonical quantization, 
  we perform two singular transformations which are dual to each other
  to quasi-particles and moving vortices respectively. 
  Just like conventional singular gauge transformation leads to
  conventional Chern-Simon (CS) term, the two mutual singular gauge
  transformations lead to mutual CS term. This elegant mutual CS term precisely
  describe the mutual AB phase scattering between vortices and quasi-particles.
  Alternatively, in path-integral presentation, the effective action describes
 the quasi-particle moving
 in both vector and scalar potentials due to the phase fluctuations of
 quantum generated vortices. By a duality transformation presented
  in Refs. \cite{dual,berry} to a vortex representation, the 
  quantum fluctuation generated vortices couple to quasi-particles
  by a mutual CS term, the vortices are also interacting with
  each other by long range logarithmic interactions due to charge fluctuation 
  described by a Maxwell term.
  By carefully considering the periodicity
  of this mutual CS term, it can be shown that
  the effective action is essentially
  equivalent to the dual vortex representation derived by Senthil and
  Fisher (SF) from the elegant $ Z_{2} $
  gauge theory \cite{z2}
  except the Volovik effect missing in the $ Z_{2} $ gauge theory
  is also explicitly incorporated into the present approach.
  Neglecting the charge fluctuation first and replacing the $ Z_{2} $
  mutual CS theory by the $ U(1) $ mutual CS theory, we find a fixed line
  characterized by the mutual statistical angle $ \theta $ and
  calculate the universal
  spinon, vortex and mutual Hall drag conductivities
  which continuously depend on $ \theta $. This transition can also be
  viewed as a simplest confinement and deconfinement transition.
  The properties of the confinement and deconfinement phases on the two sides
  of the critical point are also
  discussed. We stressed explicitly that the
  periodicity of $ Z_{2} $ mutual CS gauge theory on the lattice
  is {\em not} preserved
  by the $ U(1) $ mutual CS theory in the continuum.
  When the $ U(1) $ charge fluctuation is taken into account,
  we tentatively still replace the $ Z_{2} $ mutual CS theory by the $ U(1) $
  CS mutual theory and treat both $ U(1) $ fluctuations on the same footing.
  We find the $ U(1) $ charge fluctuation turns the fixed line into a fixed
  point where charge $ \pm e $ and spin 1/2 Dirac-like quasi-particle
  and charge $ 2e $ Cooper pairs are asymptotically decoupled.
  Although the result is pathological, it does suggest the spinon and holon
  are confined into electrons and Cooper pairs when condensing $ hc/2e $
  vortices, in contrast to the condensing of double strength $ hc/e $
  vortices discussed in \cite{balents,own} and reviewed in appendix C.
  Although we are unable to solve the critical behaviors of 
  a combined $ Z_{2} $ and $ U(1) $ theory in this paper,
  we do stress that $ Z_{2} $ periodicity should be treated correctly to
  understand the critical behaviors of a theory with both discrete $ Z_{2} $
  and continuous $ U(1) $ gauge fields. 

   The paper is organized as the following. In the next section, we derive
 the effective action with both the mutual C-S interaction and the
 charge fluctuation, first in the path integral language, and then in
 canonical quantization representation.
 In Sec. III, we concentrate on the effect of the mutual C-S interaction
 and calculate the critical exponents and universal conductivities for
 general statistical angle $ \theta $. Implications for double layers
 Quantum Hall System are given. Some
  details are relegated to appendix A. In appendix C, we concentrate on
  the effect of charge fluctuation which is the only gauge fluctuation
 in double strength $ hc/e $ vortices.
 In Sec IV, we discuss the combined effect of mutual C-S interaction and charge
 fluctuation, we also stress the importance to keep the periodicity
 of $ Z_{2} $ gauge field when considering a theory with both discrete 
 $ Z_{2} $ and continuous $ U(1) $ gauge fields. 
 In Sec. VI, we discuss the connection of our approach
 to the current fashion $ Z_{2} $ gauge theory and point out the possible
 open problems. In appendix B, we apply
 the method developed in the main text to quenched random vortex array.

\section{ The effective action}
   Following the notation in Refs.\cite{sing,static}, we define 
 $ d_{ \uparrow}= c_{\uparrow}(x), d_{ \downarrow}=
  c^{\dagger}_{\downarrow} $ \cite{exchange}. 
   Because there are equal number of positive and negative vortices,
  we divide the positive vortices into two subsets $ \phi_{p1},\phi_{p2} $
  and negative vortices into two subsets $ \phi_{n1},\phi_{n2} $.
  We define $ \phi_{A}= \phi_{p1}+\phi_{n1}, \phi_{B}= \phi_{p2}+\phi_{n2} $
  and introduce the spinon by performing
  a general singular unitary transformation $ d=Ud_{s} $:
\begin{equation}
    H_{s}=U^{-1} H U,~~~~~~~~U=
     \left ( \begin{array}{cc}
		e^{i \phi_{A}}  &   0  \\
		0  &   e^{-i \phi_{B} }  \\
		\end{array}   \right )
\label{first}
\end{equation}
 where $ \phi_{A}+ \phi_{B} = \phi $ is the phase of the Cooper pair.
  The original transformation \cite{and,sing} is devised for static vortices.
  Here we extend it to moving vortices whose phase $ \phi $
 is also fluctuating, therefore, depends on both the
 space and time. The spinon is charge neutral.
 In Anderson's gauge, $ \phi_{A}=0, \phi_{B}= \phi $ or vice versa.
 In the former(latter), the spinon is electron-like ( hole-like).

   Expanding $ H_{s} $ around the node 1 where $ \vec{p}=(p_{F},0) $,
   we obtain the linearized quasi-particle Lagrangian $ {\cal L}_{qp} $
   in the presence of the external gauge potential $ A_{\mu} $:
\begin{eqnarray}
   {\cal L}^{u}_{qp} &= & \psi^{\dagger}_{1} [ (\partial_{\tau} -i a_{\tau} )
    + v_{f} (p_{x}-  a_{x} ) \tau^{3} + v_{2} (p_{y}-  a_{y} ) \tau^{1} ]
    \psi_{1}    \nonumber   \\
  & + & \psi^{\dagger}_{1} \psi_{1} v_{f} v_{x}(\vec{r}) 
    + i \psi^{\dagger}_{1} \tau^{3} \psi_{1} v_{\tau}(\vec{r}) 
    + (1 \rightarrow 2, x \rightarrow y )
\label{linear}
\end{eqnarray}
   where $ v_{\mu}= \frac{\hbar}{2} \partial_{\mu} \phi-\frac{e}{c} 
   A_{\mu} $ is the gauge-invariant superfluid {\em momentum}, it acts
   as a scalar scattering potential,
   $ a_{\mu}= \frac{1}{2} \partial_{\mu} ( \phi_{A} -\phi_{B} ) $ is the AB
   gauge field due to the phase winding of vortices \cite{curve}.
  Note that the external gauge potential only appear explicitly in the
  superfluid momentum $ v_{\mu} $.
   Because there are equal number of positive and negative vortices in both
   $ \phi_{A} $ and $ \phi_{B} $,
   on the average, the vanishing of $ v_{\mu} $ and $ a_{\mu} $ is
   automatically ensured, in this case, the Anderson gauge
   turns out to be more convenient than FT gauge.
    In the following, we use the electron-like Anderson gauge
    where $ a_{\mu}= \frac{1}{2} \partial_{\mu} \phi $
    \cite{hole}.

    We get the corresponding expression at node $ \bar{1} $ and $ \bar{2} $
   by changing
   $ v_{f} \rightarrow -v_{f}, v_{2} \rightarrow -v_{2} $ in the above Eq.
\begin{eqnarray}
   {\cal L}^{l}_{qp} &= & \psi^{\dagger}_{\bar{1}} [ (\partial_{\tau} -i a_{\tau} )
    - v_{f} (p_{x}-  a_{x} ) \tau^{3} - v_{2} (p_{y}-  a_{y} ) \tau^{1} ]
    \psi_{\bar{1}}    \nonumber   \\
  & - & \psi^{\dagger}_{\bar{1}} \psi_{\bar{1}} v_{f} v_{x}(\vec{r}) 
    + i \psi^{\dagger}_{\bar{1}} \tau^{3} \psi_{\bar{1}} v_{\tau}(\vec{r}) 
    + (\bar{1} \rightarrow \bar{2}, x \rightarrow y )
\end{eqnarray}

   Performing a P-H transformation
   $ \tilde{\psi}_{1\alpha}= \epsilon_{\alpha \beta}
   \psi^{\dagger}_{\bar{1}\beta} $ and the
   corresponding expression at node $ \bar{2} $, the above Eq. becomes:
\begin{eqnarray}
   {\cal L}^{l}_{qp} &= & \tilde{\psi}^{\dagger}_{1} [ (\partial_{\tau} +
   i a_{\tau} )
    + v_{f} (p_{x}+  a_{x} ) \tau^{3} + v_{2} (p_{y}+  a_{y} ) \tau^{1} ]
    \tilde{\psi}_{1}    \nonumber   \\
  & + & \tilde{\psi}^{\dagger}_{1} \tilde{\psi}_{1} v_{f} v_{x}(\vec{r}) 
  + i \tilde{\psi}^{\dagger}_{1} \tau^{3} \tilde{\psi}_{1} v_{\tau}(\vec{r}) 
  + (1 \rightarrow 2, x \rightarrow y )
\label{tilde}
\end{eqnarray}

   In order to make the final expressions explicitly $ SU(2) $ invariant,
   we perform the singular gauge transformation 
   $ \psi_{12\alpha}= e^{-i(\phi_{A}-\phi_{B} )}
    \tilde{\psi}_{1 \alpha} $ and the
  corresponding transformation at node $ \bar{2} $, then $ a_{\mu}
  \rightarrow -a_{\mu} $, Eq.\ref{tilde} takes
  the same form as Eq.\ref{linear} \cite{break}. Adding the two equations
   leads to: 
\begin{eqnarray}
   {\cal L}_{qp} &= & \psi^{\dagger}_{1a} [ (\partial_{\tau} -i a_{\tau} )
    + v_{f} (p_{x}-  a_{x} ) \tau^{3} + v_{2} (p_{y}-  a_{y} ) \tau^{1} ]
    \psi_{1a}    \nonumber   \\
  & + & \psi^{\dagger}_{1a} \psi_{1a} v_{f} v_{x}(\vec{r}) 
    + i \psi^{\dagger}_{1a} \tau^{3} \psi_{1a} v_{\tau}(\vec{r}) 
    + (1 \rightarrow 2, x \rightarrow y )
\label{linear4}
\end{eqnarray}
  where $ a=1,2 $ is the spin indices. $ \tau^{\prime s} $ matrices are acting
  on {\em particle-hole} space.
   As intended, the above Eq. is explicitly $ SU(2) $ spin invariant.
  Equivalently, we can start with the explicitly spin $ SU(2) $ 
  invariant approach advocated in Ref.\cite{balents} and perform the 
  singular gauge transformation in the {\em p-h } space. 

 $ {\cal L}_{qp} $ enjoys gauge symmetry $ U(1)_{u}
  \times U_{s}(1) $ ( in fact, it is $ U(1) \times Z_{2} $ ),
  the first being uniform and second being staggered 
  gauge symmetry:
\underline{Uniform (or external) $ U_{u}(1) $ gauge symmetry }
\begin{eqnarray}
  c_{\alpha} & \rightarrow & c_{\alpha} e^{i \chi},~~~ 
  d_{\alpha} \rightarrow d_{\alpha}   \nonumber \\
  \phi_{A} & \rightarrow & \phi_{A} + \chi ,~~~~
  \phi_{B} \rightarrow \phi_{B} + \chi
\label{ext}
\end{eqnarray}
   Under this uniform $ U(1) $ transformation, the corresponding fields
   transform as:
\begin{eqnarray}
  \phi & \rightarrow & \phi + 2\chi ,~~~~ A_{\alpha} \rightarrow 
  A_{\alpha} + \partial_{\alpha} \chi
	     \nonumber   \\
  v_{\alpha} & \rightarrow & v_{\alpha},~~~~~a_{\alpha} \rightarrow a_{\alpha}
\end{eqnarray}

      $ d, v_{\alpha}, a_{\alpha} $ all are invariant under this external
      $ U(1) $ transformation. Therefore the spinon $ d_{\alpha} $
      is charge {\em neutral} to the external magnetic field.

\underline{Staggered (or internal) $ U_{s}(1) $ gauge symmetry}
\begin{eqnarray}
  c_{\alpha} & \rightarrow & c_{\alpha},~~~ 
  d_{\alpha} \rightarrow d_{\alpha} e^{-i \chi}   \nonumber \\
  \phi_{A} & \rightarrow & \phi_{A} + \chi ,~~~~
  \phi_{B} \rightarrow \phi_{B} - \chi
\label{int}
\end{eqnarray}
   Under this internal $ U(1) $ transformation, the corresponding fields
   transform as:
\begin{eqnarray}
  \phi  & \rightarrow & \phi ,~~~~ A_{\alpha} \rightarrow A_{\alpha} 
	     \nonumber   \\
   v_{\alpha} & \rightarrow & v_{\alpha},~~~~~a_{\alpha} \rightarrow a_{\alpha}+
      \partial_{\alpha} \chi
\end{eqnarray}

      Although the spinon $ d_{\alpha} $
      is charge neutral to the external magnetic field, it carries
      charge $ 1 $ to the internal gauge field $ a_{\alpha} $.

   It is easy to realize that $ U_{u}(1) $ acts only on the boson sector,
  since the spinon is charge neutral, $ U_{s}(1) $ acts only on the
   fermion sector. In fact, $ U_{s}(1) $ should
   be a discrete local $ Z_{2} $ symmetry \cite{z2}, because up and down
  $ hc/2e $ vortices are equivalent and do not break T symmetry \cite{static}.

     The phase fluctuation is simply $ 2+1 $ dimensional X-Y model:
\begin{equation}
    {\cal L}_{ph}= \frac{K}{2} v^{2}_{\mu} =
    \frac{K}{2} ( \partial_{\mu} \phi -2 A_{\mu} )^{2}
\end{equation}
  
    After absorbing the scalar potential scattering part ( Volovik term)
   into $ {\cal L}_{ph} $,   we can write the total Lagrangian
   $ {\cal L}= {\cal L}_{qp}+ {\cal L}_{ph} $ as:
\begin{eqnarray}
   {\cal L} &= & \psi^{\dagger}_{1a} [ (\partial_{\tau} -i a_{\tau} )
    + v_{f} (p_{x}-  a_{x} ) \tau^{3} + v_{2} (p_{y}-  a_{y} ) \tau^{1} ]
    \psi_{1a}    \nonumber   \\
    & + &  (1 \rightarrow 2, x \rightarrow y )
      + \frac{K}{2} ( \partial_{\mu} \phi - A^{eff}_{\mu} )^{2}
\label{act}
\end{eqnarray}
   where $ A^{eff}_{\mu}= 2 A_{\mu} -K^{-1} J_{\mu} $ and the quasi-particle
   electric current is : $ J_{0} = \psi^{\dagger}_{j} \tau^{3}
   \psi_{j}, J_{x} = v_{f} \psi^{\dagger}_{1} \psi_{1},
   J_{y} =v_{f} \psi^{\dagger}_{2} \psi_{2} $.

    From Eq.\ref{act}, it is easy to identify the two conserved Noether
    currents: spinon current and electric current.

    The spinon current is given by:
\begin{eqnarray}
   j^{s}_{0} & = & \psi^{\dagger}_{1}(x) \psi_{1}(x) + \psi^{\dagger}_{2}(x) 
                \psi_{2}(x)
                       \nonumber  \\
   j^{s}_{x} & = & \psi^{\dagger}_{1}(x)  v_{F} \tau^{3} \psi_{1}(x)
  + \psi^{\dagger}_{2}(x)  v_{2} \tau^{1} \psi_{2}(x)
                       \nonumber  \\
  j^{s}_{y} & = & \psi^{\dagger}_{1}(x)  v_{2} \tau^{1} \psi_{1}(x)
   + \psi^{\dagger}_{2}(x)  v_{F} \tau^{3} \psi_{2}(x)
\label{gap}
\end{eqnarray}

    Obviously the spinon current only comes from quasi-particle.
  In principle, the spinon current is not conserved due to scatterings between
  different nodes which lead to anomalous terms not included in Eq.\ref{act}.
  However, the inter-node scatterings involve large momenta transfer
  $ \vec{K}_{i} -\vec{K}_{j} $ for $ i \neq j $,
  so we neglect them due to momentum conservation
 in the long wave-length limit of the phase fluctuation \cite{magnetic}.
   The exact conserved {\em spin current}
  $ \vec{j}^{S}_{\mu} $ is with $ \vec{\sigma}/2 $ inserted in the above
  spinon currents.

     The electric current is given by:
\begin{equation}
    j^{e}_{\mu}= -\frac{\partial {\cal L}}{ \partial A^{eff}_{\mu} }
                  = K (\partial_{\mu} \phi - A^{eff}_{\mu})
                  = K (\partial_{\mu} \phi - 2 A_{\mu}) + J_{\mu}
\end{equation}
 Where the first part coming from Cooper pair and the second from
 the quasi-particle. Although they are not separately conserved, their sum is.
    
   The Noether current due to the the symmetry under $ \phi \rightarrow \phi
   + \chi $ can be written as:
\begin{equation}
    j^{t}_{\mu}= \frac{\partial {\cal L}}{ \partial (\partial_{\mu} \phi)}
                  = K (\partial_{\mu} \phi - A^{eff}_{\mu}) 
                    - \frac{1}{2} j^{s}_{\mu}
                  =  j^{e}_{\mu} -  \frac{1}{2} j^{s}_{\mu}
\end{equation}
    
     It is a combination of electric and spinon currents, therefore also
   conserved.

    Following Refs.\cite{dual,berry}, we perform a duality transformation
   to Eq.\ref{act}
\begin{eqnarray}
   & &    \frac{K}{2} ( \partial_{\mu} \phi - A^{eff}_{\mu} )^{2} 
   -i \frac{1}{2} \partial_{\mu} \phi j^{s}_{\mu}
            \nonumber  \\
   &= & i j_{e \mu} ( \partial_{\mu} \phi - A^{eff}_{\mu} ) + \frac{1}{2 K}
     j^{2}_{ e \mu} -i \frac{1}{2} \partial_{\mu} \phi j^{s}_{\mu}
                      \nonumber  \\
   &= & i j^{t}_{\mu} \partial_{\mu} \theta + 
    i  j^{t}_{ \mu} \partial_{\mu} \phi 
    - i j^{e}_{\mu} A^{eff}_{\mu}  +  \frac{1}{2 K} j^{2}_{ e \mu}
\label{boson}
\end{eqnarray}
   Where we have separated topological trivial spin-wave part and
   topological non-trivial vortex parts.

    Integrating out the spin-wave part, we get the conservation equation for
  the total current $ j^{t}_{\mu}= j^{e}_{\mu} - 
  \frac{1}{2} j^{s}_{\mu} $. In fact, as
  shown in the previous paragraphs, $ j^{e}_{\mu} $ and $ j^{s}_{\mu} $ are
  separately conserved. Therefore we can introduce spin and electric gauge
  fields by:
\begin{eqnarray}
     j^{s}_{\mu} & = & \epsilon_{\mu \nu \lambda } \partial_{\nu} a^{s}_{\lambda}
               \nonumber  \\
     j^{e}_{\mu} &= & \epsilon_{\mu \nu \lambda } \partial_{\nu} a^{e}_{\lambda}
\end{eqnarray}

      We can also define the vortex current:
\begin{eqnarray}
     j^{v}_{\mu} & = & \frac{1}{2 \pi} 
    \epsilon_{\mu \nu \lambda } \partial_{\nu} \partial_{\lambda} \phi 
                \nonumber  \\
     & = & \epsilon_{\mu \nu \lambda } \partial_{\nu} a^{v}_{\lambda}
\end{eqnarray}
      where $ a^{v}_{\mu} = \partial_{\mu} \phi/ 2 \pi $ is the vortex gauge
   field.
   
     Substituting the above expressions into Eq.\ref{boson}, we reach:
\begin{eqnarray}
  & &  \frac{1}{ 4 K} f^{2}_{e \mu \nu} + i \partial_{\mu} \phi j^{t}_{\mu}
   -i A^{eff}_{\mu} \epsilon_{\mu \nu \lambda } \partial_{\nu} a^{e}_{\lambda}
            \nonumber  \\
  & = & \frac{1}{ 4 K} f^{2}_{e \mu \nu} + i a^{t}_{\mu} j^{v}_{\mu}
   -i A^{eff}_{\mu} \epsilon_{\mu \nu \lambda } \partial_{\nu} a^{e}_{\lambda}
\end{eqnarray}
    where $ a^{t}_{\mu}= a^{e}_{\mu}- \frac{1}{2} a^{s}_{\mu} $ is the total
  gauge field felt by the moving vortices.

    Using $ \Phi $ for the vortex operator and adding the 
  quasi-particle part, we get the following effective action:
\begin{eqnarray}
   {\cal L} &= & \psi^{\dagger}_{1} [ (\partial_{\tau} -i a^{\psi}_{\tau} )
    + v_{f} (p_{x}-  a^{\psi}_{x} ) \tau^{3} + v_{2} (p_{y}-  a^{\psi}_{y} )
    \tau^{1} ]
    \psi_{1}   \nonumber  \\
       &  + &  (1 \rightarrow 2, x \rightarrow y )  
                       \nonumber  \\
  & + & | ( \partial_{\mu} -i a^{\Phi}_{\mu}-i a_{\mu} ) \Phi |^{2}
      + V( |\Phi|) + \frac{i}{ 2 \pi \theta} a^{\psi}_{\mu} 
    \epsilon_{\mu \nu \lambda } \partial_{\nu} a^{\Phi}_{\lambda}
                        \nonumber  \\
  & + & \frac{1}{ 4 } f^{2}_{ \mu \nu} 
   -i A^{eff}_{\mu} \epsilon_{\mu \nu \lambda } \partial_{\nu} a_{\lambda}
    - \mu \epsilon_{ij} \partial_{i} a_{j}
\label{dual}
\end{eqnarray}
  where $ V(|\Phi|)= m^{2}_{\Phi} |\Phi|^{2}+ g_{\Phi} |\Phi|^{4} +\cdots $
  stands for the short range interaction between the vortices.
  The last term is due to Berry phase in the boson representation \cite{berry}
   which can be absorbed into $ A^{eff}_{\mu} $ by redefining 
   $ A^{eff}_{\mu} \rightarrow A^{eff}_{\mu} +i \mu \delta_{\mu 0} $,
   it acts like an external magnetic field in the $ \hat{z} $
  ( namely $\mu=0 $ ) direction.
  There are one species of vortex and $ N=4 $ species of Dirac
  fermion( 2 spin components at 2 nodes at the upper plane),
  the mutual statistical angle $ \theta= \pm 1/2 $. We also changed
  the notation by setting $ a^{\psi} = a^{v}, a^{\Phi}= a^{s}, a^{e}=a $.

    Although we derived the above equation in the electron-like Anderson
    gauge, in fact, it holds in {\em any} gauge.
    Eq.\ref{dual} again enjoys the gauge symmetry $ U(1) \times U(1) $,
   the first only acting on the vortex sector is the electric $ U(1) $
   gauge field, the second on both
   the vortex and fermion sectors is the $ U(1) $ mutual CS gauge field.
    The mutual Chern-Simon term enforces the constraints:
   $ 2 \pi \theta j^{s}_{\mu}  =  \epsilon_{\mu \nu \lambda } \partial_{\nu} 
    a^{\Phi}_{\lambda}, 2 \pi \theta j^{v}_{\mu}  =  \epsilon_{\mu \nu \lambda }
    \partial_{\nu} a^{\psi}_{\lambda} $. Physically, it means that when 
    a quasi-particle encircles
    a vortex, it picks up a phase $ 2 \pi \theta $. Equivalently
    when a vortex moves around a quasi-particle, it also picks up a phase
    $ 2 \pi \theta $.  Although the conventional C-S term breaks T symmetry
     and has  periodicity under $ \theta \rightarrow \theta +2 $
     \cite{hlr,chen,wen,boson,subir}, the
    mutual C-S term does {\em not } break T symmetry and has 
     periodicity under $ \theta \rightarrow \theta +1 $. For example,
     $ \theta=-1/2 $ is equivalent to $ \theta=1/2 $. In fact, this mutual
     CS term at $ \theta=1/2 $ should be the same as the $ Z_{2} $
     mutual CS term on the lattice discussed in Ref.\cite{z2}.
     Therefore strictly speaking $ a^{\psi} $ and $ a^{\Phi} $ are $ Z_{2} $
     gauge fields, only $ a_{\mu} $ is a  $ U(1) $ gauge field.
     The Doppler-shifted term ( Volovik effect) is encoded in the last term
     in Eq.\ref{dual}. This effect was {\em not} taken into account in
     the $ Z_{2} $ gauge theory \cite{z2}.

   In canonical quantization language, it is very instructive to compare
   the well-known singular gauge transformation leading to composite fermion
   in $ \nu=1/2 $ system by Halperin, Lee and Read (HLR) \cite{hlr} to
   the singular gauge transformations performed in the spinon-vortex system
   in this paper.  The crucial difference  is that in the former,
    we attach electron's own $ \theta $
  flux to itself ( $ \theta =2 $ in $ \nu=1/2 $ system to keep fermion
  statistics intact ), the constraint  $ \nabla \times \vec{a}(\vec{r})
  = 2 \pi \theta \rho(\vec{r}) $  in Coulomb gauge
  $ \partial_{i} a_{i} =0 $
  leads to composite fermion $ \psi_{c} $ coupled
  to conventional CS term which breaks T symmetry, has periodicity
  under $ \theta \rightarrow \theta +2 $; however,
  in the latter, we attach vortex's
  $ \theta=1/2 $ flux to quasi-particle or {\em vise versa},
  the two constraints $ \nabla \times \vec{a}^{\psi}(\vec{r})
  = 2 \pi \theta j^{v}_{0}(\vec{r}), \nabla \times \vec{a}^{\Phi}(\vec{r})
  = 2 \pi \theta j^{s}_{0}(\vec{r}) $ in Coulomb gauges
  $ \partial_{i} a^{\psi}_{i} = \partial_{i} a^{\Phi}_{i} =0 $
   naturally leads to quasi-particles and vortices are
  coupled by mutual CS term in additional to the electrical gauge
  field which mediates the logarithmic interaction between vortices.
  The mutual CS term does
  not break T symmetry, has periodicity under $ \theta \rightarrow \theta +1 $.

   It is important to point out that when a vortex  is moving around a 
  closed loop, it pick up two phases, one is $ Z_{2} $ phase
  due to the spinon current, described
  by the mutual C-S term, another is $ U(1) $ phase
  due to the total electric charge current
  described by the Maxwell term. In the following section, we will neglect
  the electric charge fluctuation and concentrate the mutual statistics term.
  In Sec. IV, we will study the effect of charge fluctuation described by the
  Maxwell term. For completeness, we relegate the review of the effect of charge
  fluctuation to appendix B.
   
     For simplicity, we take the relativistic form for both fermion and boson,
   because the anisotropies in Eq.\ref{dual} are expected to be irrelevant
   near the zero temperature QCP \cite{guess}.

\section{ The effect of mutual statistics }
    In this section, we neglect the charge fluctuation, namely setting
    $ a_{\mu}=0 $ in Eq.\ref{dual}.
    The charge fluctuation can be suppressed by condensing $ hc/e $ vortices
    reviewed in appendix C.
    We could add the kinetic and potential terms for the $ hc/e $ vortex
    operator $ \Phi_{2} $ to Eq.\ref{dual}:
\begin{equation}
  {\cal L}_{\Phi_{2}}= | ( \partial_{\mu} -i 2 a_{\mu} ) \Phi_{2} |^{2}
   + V( |\Phi_{2}|) 
\end{equation}
     As shown in appendix C, there is no mutual statistical interaction
   between spinon and $ hc/e $ vortices $ \Phi_{2} $. The long range
   logarithmic interaction between $ hc/2e $ vortices $ \Phi $ and $ hc/e $
   vortices $ \Phi_{2} $ is mediated by the electrical gauge field $ a_{\mu} $.
   Condensing $ < \Phi_{2} > = \Phi_{20} $ will generate a mass term
   $ \frac{\Phi^{2}_{20}}{2} (a_{\mu})_{t}^{2} $ which dominates 
   over the Maxwell term. Integrating out $ a_{\mu} $ leads to
\begin{equation}
   \frac{1}{\Phi^{2}_{20} }[ \frac{1}{ 4} ( f^{A^{eff}}_{\mu \nu} )^{2} 
   -i  A^{eff}_{\mu} \epsilon_{\mu \nu \lambda } \partial_{\nu} 
   j^{v}_{\lambda} + (j^{v}_{\mu})^{2}_{t}  ]
\end{equation}
  
 All the generated terms  only renormalize
   the short range interactions already included in $ V(|\Phi| ) $.
    The model is also closely related to the Hall drag problem
   in double layer Quantum Hall system \cite{kun}, therefore  
   the model itself is interesting on its own right and deserves detailed
   investigation. 
\subsection{ Quantum Critical point}
   In order to calculate conductivities, we add two source fields
   $ A^{\psi}_{\mu} $ and $ A^{\Phi}_{\mu} $ for the quasi-particles and
    vortex respectively:

\begin{eqnarray}
   {\cal L} &= & \psi^{\dagger}_{a} \gamma_{\mu} (\partial_{\mu}
    -i a^{\psi}_{\mu} -i A^{\psi}_{\mu} ) \psi_{a}   \nonumber  \\
  & + & | ( \partial_{\mu} -i a^{\Phi}_{\mu}-i A^{\Phi}_{\mu} ) \Phi |^{2}
      + V( |\Phi|) + \frac{i}{ 2 \pi \theta} a^{\psi}_{\mu} 
    \epsilon_{\mu \nu \lambda } \partial_{\nu} a^{\Phi}_{\lambda}
\label{mutual}
\end{eqnarray}
    where $ a=1,2,3,4 $ stands for $ N=4 $ species of Dirac fermion.
   In fact, two Maxwell terms for $ a^{\psi} $ and $ a^{\Phi} $ can be
   added to the above equation, but they are subleading to the mutual CS
   term in the low energy limit. It is expected that there is no periodicity
   under $ \theta \rightarrow \theta +1 $ in the continuum limit.

   If the theory is properly regularized on the lattice,
  it should has the periodicity
  under $ \theta \rightarrow \theta +1 $, for example, $ \theta= -1/2
  $ should be equivalent to $ \theta =1/2 $. In fact, $ \theta= 1/2 $
  mutual $ U(1) $
  CS theory could be the same as the $ Z_{2} $ mutual CS theory
  developed in \cite{z2}.

    The RG calculation in Refs.\cite{chen,wen,boson} can be used to show
   that $ \theta $
   is exactly marginal, therefore there is a line of fixed points determined
   by the mutual statistical angle $ \theta $.
   In order to calculate the {\em spin conductivity} along this fixed line,
   a source field could be introduced to couple to  the spin current
   $ \vec{j}^{S}_{\mu}=
   \psi^{\dagger}_{\alpha}
    \gamma_{\mu} (\vec{\sigma})_{\alpha \beta} \psi_{\beta} $.
   Similar calculations follow.

     Integrating out both fermion and boson leads to:
\begin{eqnarray}
   {\cal L} & = & -\frac{1}{2} a^{\psi}_{\mu} (-k)
         \Pi^{\psi}_{\mu \nu}(k) a^{\psi}_{\nu}(k)
    -\frac{1}{2} a^{\psi}_{\mu} (-k)
         \Pi^{\psi \Phi}_{\mu \nu}(k) a^{\Phi}_{\nu}(k)          
                     \nonumber  \\
   & - & \frac{1}{2} a^{\Phi}_{\mu} (-k)
         \Pi^{\Phi}_{\mu \nu}(k) a^{\Phi}_{\nu}(k)
    -\frac{1}{2} a^{\Phi}_{\mu} (-k)
         \Pi^{\Phi \psi}_{\mu \nu}(k) a^{\psi}_{\nu}(k)
                      \nonumber  \\
   & - & \frac{1}{2 \theta} ( a^{\psi}_{\mu} (-k) - A^{\psi}_{\mu} (-k)) 
         \epsilon_{\mu \nu \lambda} k_{\lambda} ( a^{\Phi}_{\nu}(k)
         - A^{\Phi}_{\nu}(k))
                         \nonumber  \\
   & - &  \frac{1}{2 \theta} ( a^{\Phi}_{\mu} (-k) - A^{\Phi}_{\mu} (-k)) 
         \epsilon_{\mu \nu \lambda} k_{\lambda} ( a^{\psi}_{\nu}(k)
         - A^{\psi}_{\nu}(k))
\label{gen}
\end{eqnarray}
    where the exact forms of $ \Pi^{\prime s} $ are dictated by gauge
   invariance and Furry's theorem :
\begin{eqnarray}
 \Pi^{\psi}_{\mu \nu}(k) &= & \Pi_{1}(k) k ( \delta_{\mu \nu} -
    \frac{ k_{\mu} k_{\nu} }{k^{2}} )   \nonumber  \\
 \Pi^{\psi \Phi}_{\mu \nu}(k) &= & \Pi^{\Phi \psi}_{\mu \nu}(k) = 
 \Pi_{2}(k) \epsilon_{\mu \nu \lambda} k_{\lambda} 
                       \nonumber  \\
 \Pi^{\Phi}_{\mu \nu}(k) &= & \Pi_{3}(k) k ( \delta_{\mu \nu} -
    \frac{ k_{\mu} k_{\nu} }{k^{2}} )
\end{eqnarray}

   Where $ \Pi_{1}, \Pi_{2}, \Pi_{3} $ are the polarizations for
  $ a^{\psi} a^{\psi}, a^{\psi} a^{\Phi}, a^{\Phi} a^{\Phi} $.

  If we are only interested in DC conductivities, for simplicity, we can
  put $ \vec{k} =0 $, Eq.\ref{gen} becomes ( for the most general form,
    see appendix A): 
\begin{eqnarray}
   {\cal L} & = & -\frac{1}{2} a^{\psi}_{i} (-\omega_{n})
         \Pi_{1} | \omega_{n} | a^{\psi}_{i}(\omega_{n})
    -\frac{1}{2} a^{\psi}_{i} (-\omega_{n})
         \Pi_{2} \epsilon_{ij} \omega_{n} a^{\Phi}_{j}(\omega_{n})          
                     \nonumber  \\
    & - & \frac{1}{2} a^{\Phi}_{i} (-\omega_{n})
         \Pi_{3} | \omega_{n} | a^{\Phi}_{i}(\omega_{n})
    -\frac{1}{2} a^{\Phi}_{i} (-\omega_{n})
         \Pi_{2} \epsilon_{ij} \omega_{n} a^{\psi}_{j}(\omega_{n})          
                      \nonumber  \\
   & - & \frac{1}{2 \theta} ( a^{\psi}_{i} (-\omega_{n}) 
    - A^{\psi}_{i} (-\omega_{n})) 
         \epsilon_{ij} \omega_{n} ( a^{\Phi}_{j}(\omega_{n})
         - A^{\Phi}_{j}(\omega_{n}))
                         \nonumber  \\
   & - & \frac{1}{2 \theta} ( a^{\Phi}_{i} (-\omega_{n}) 
    - A^{\Phi}_{i} (-\omega_{n})) 
         \epsilon_{ij} \omega_{n} ( a^{\psi}_{j}(\omega_{n})
         - A^{\psi}_{j}(\omega_{n}))
\end{eqnarray}

    If we define $ \tilde{a}^{\Phi}_{i} (\omega_{n})
    =\epsilon_{ij} a^{\Phi}_{j} (\omega_{n}),
     \tilde{A}^{\Phi}_{i} (\omega_{n}) 
    =\epsilon_{ij} A^{\Phi}_{j} (\omega_{n}) $,
     the above
    equation becomes diagonal in the spatial indices $ i=1,2 $.
    Finally, integrating out $ a^{\psi}_{i}, \tilde{a}^{\Phi}_{i} $ leads to
\begin{equation}
     {\cal L} = -\frac{1}{2} ( A^{\psi}_{i}, \tilde{A}^{\Phi}_{i} )
                                \left( \begin{array}{cc}
				\sigma^{\psi} &   \sigma^{H}  \\
				\sigma^{H}  &   \sigma^{\Phi}  \\
				\end{array}   \right )
                                \left( \begin{array}{c}
				A^{\psi}_{i}   \\
				\tilde{A}^{\Phi}_{i}  \\
				\end{array}   \right )
\label{matrix}
\end{equation}
   
   Where spinon, vortex and mutual Hall drag conductivities are:
\begin{eqnarray}
   \sigma^{\psi} & = & ( \frac{1}{\theta} )^{2} \frac{\Pi_{1}}{
    \Pi_{1} \Pi_{3} + (1/\theta -\Pi_{2})^{2} }  \nonumber  \\
   \sigma^{\Phi} & = & ( \frac{1}{\theta} )^{2} \frac{\Pi_{3}}{
    \Pi_{1} \Pi_{3} + (1/\theta -\Pi_{2})^{2} }  \nonumber  \\
   \sigma^{H} & = & ( \frac{1}{\theta} )^{2} \frac{\Pi_{2} -\theta ( \Pi_{1}
     \Pi_{3} + \Pi^{2}_{2} )}{
    \Pi_{1} \Pi_{3} + (1/\theta -\Pi_{2})^{2} }
\label{three}
\end{eqnarray}

      In fact, all the three conductivities can be written in the elegant
    connection formula:
\begin{equation}
   \rho_{ij}= (\rho_{FB} )_{ij} - \theta \epsilon_{ij}
\end{equation}
     with the conductivity tensor of fermion and boson given by
\begin{equation}
                      \sigma_{FB}=  \left( \begin{array}{cc}
				\Pi_{1} &  - \Pi_{2}  \\
				\Pi_{2}  &   \Pi_{3}  \\
				\end{array}   \right )
\label{exact}
\end{equation}

  Although its form is similar to
 the conventional connection formulas discussed in \cite{hlr,chen,wen},
 the physical interpretations of the conductivities are quite different
 (see the following).

   When the vortices are generated by external magnetic field and pinned
  by impurities as discussed in \cite{static}, the total conductivity is
  the same as the fermion conductivity because the static vortex do not
  contribute. As explained in appendix B, the static vortices only feel
  the electric gauge field $ a_{\mu} $, but {\em not}
  the statistical gauge field $ a^{\Phi}_{\mu} $.

 The above expressions are exact, but $ \Pi_{1}, \Pi_{2}, \Pi_{3} $
 can only be calculated perturbatively in the coupling constant
  $ g^{2}= 2 \pi \theta $. The renormalized propagators for $ a^{\psi} $ and $ a^{\Phi} $
  can be found from the following Feymann diagrams:

\vspace{0.25cm}

\epsfig{file=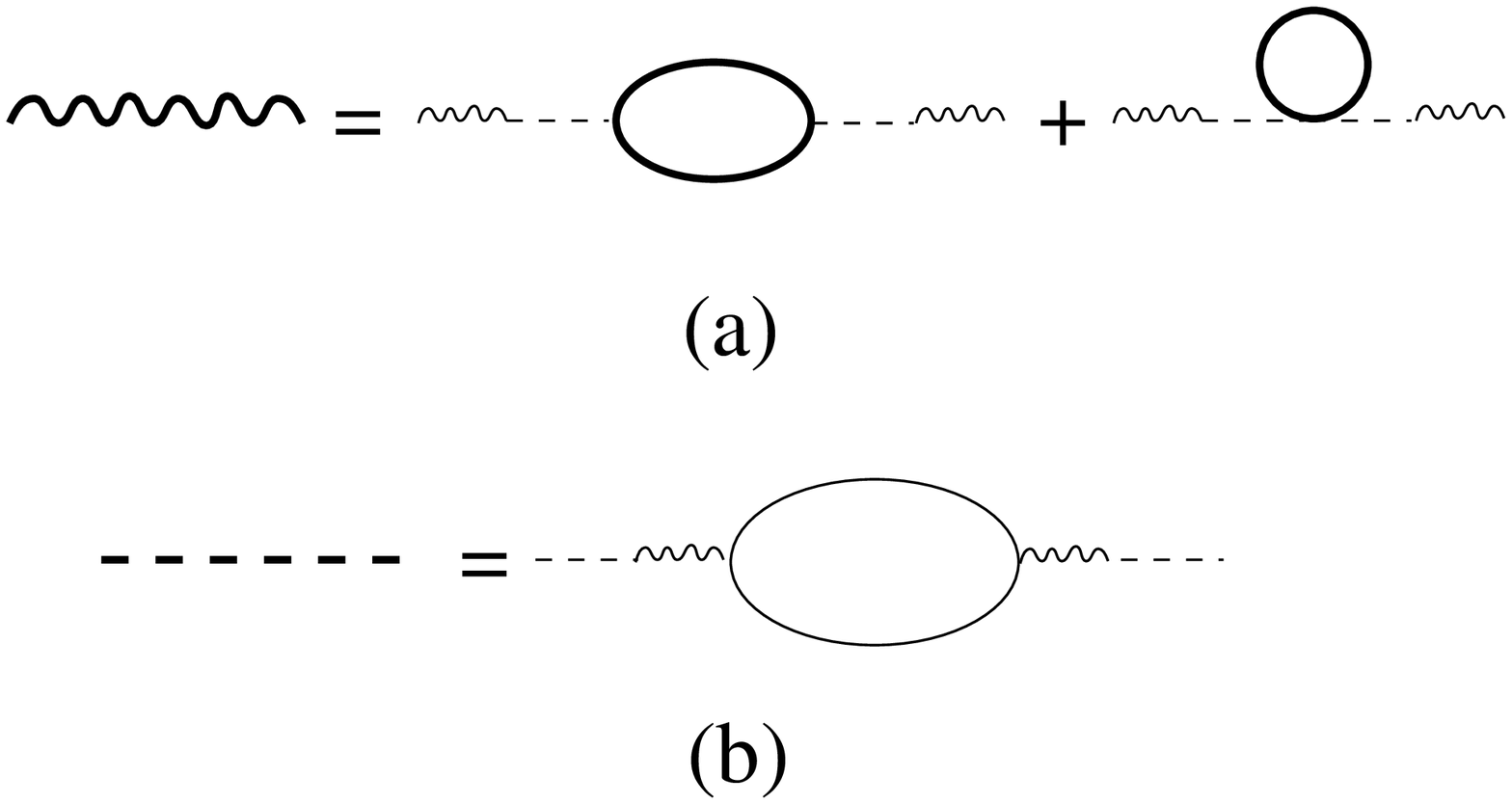,width=3.2in,height=0.9in,angle=0}

{\footnotesize {\bf Fig 2:} The renormalized propagators for $ a^{\psi} $ (thick wiggle line )
    and $ a^{\Phi} $ (thick dashed line). The thin wiggle line stands for $ a^{\psi} $,
   the dashed thin line for $ a^{\Phi} $, the thin solid line for fermion
    propagators, the thick solid line for the boson propagators.}

\vspace{0.25cm}

  By using the bare propagators $ <a^{\psi}_{\mu} a^{\Phi}_{\nu} > =-\epsilon_{\mu \nu \lambda}
     k_{\lambda}/k^{2} $ and the bare fermion and boson loop results
   $ \Pi^{f0}_{\mu \nu} = \Pi^{b0}_{\mu \nu} = -\frac{g^{2}}{16}
   k ( \delta_{\mu \nu} -k_{\mu} k_{\nu}/k^{2} ) $, we can find easily the
   renormalized $ a^{\psi} $ and $ a^{\Phi} $ propagators
    $ G^{\psi}_{\mu \nu} = G^{\Phi}_{\mu \nu} = -\frac{g^{2}}{16} \frac{1}{k}
    ( \delta_{\mu \nu} -k_{\mu} k_{\nu}/k^{2} ) $. In contrast to the conventional
   CS theory studied in \cite{chen,wen}, the propagators are {\em even} in $ k $,
   this is because the theory respects $ T $ symmetry. On the other hand, in contrast to
   the Maxwell propagators, they behave as $ 1/k $ instead of $ 1/k^{2} $. 

    The three loop diagrams for $ \Pi_{1} $ are given by:

\vspace{0.25cm}

\epsfig{file=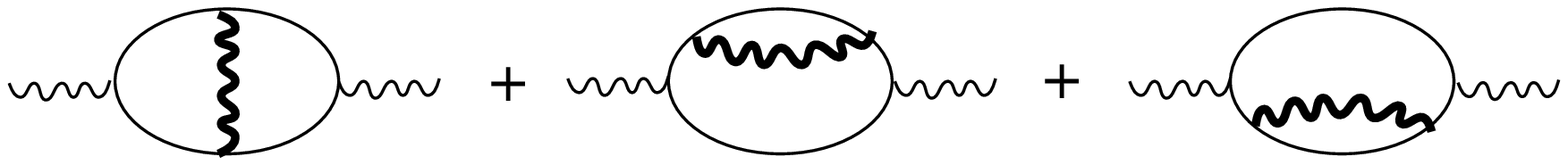,width=3.2in,height=0.6in,angle=0}

{\footnotesize {\bf Fig 3:} The three loop diagrams of $ \Pi_{1} $. The thick wiggle line
  stands for the renormalized propagators of $ a^{\psi} $,
  the thin solid line stands for the fermion propagator.  The one loop diagram is not shown.}    

\vspace{0.25cm}

   By using $ G^{\psi}_{\mu \nu} $ and
   extracting the symmetric part of the gauge propagator in the large
   $ N $ results in Refs.\cite{chen,wen}, we are able to calculate the above three
   loop diagrams.  Furry's theorem can be used to eliminate large number of null diagrams.
   We get the following series:
\begin{equation}
  \Pi_{1} = N\frac{\pi}{8}
  (1+ \frac{3}{16} \frac{g^{4}}{ (2 \pi)^{2} } +  g^{8} + \cdots)
\end{equation}
  where $ N=4 $ is due to the sum over 4 species of Dirac fermions.

    The three loop diagrams for $ \Pi_{3} $ are given by:

\vspace{0.25cm}

\epsfig{file=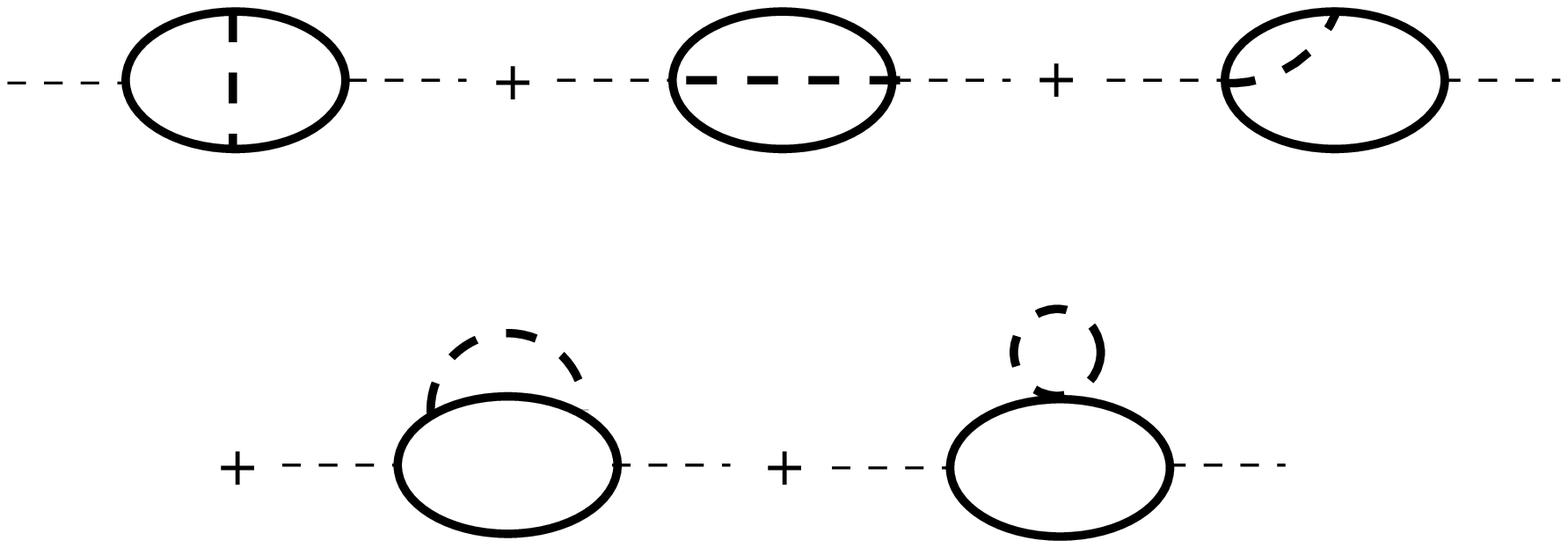,width=3.2in,height=1in,angle=0}

{\footnotesize {\bf Fig 4:} The three loop diagrams of $ \Pi_{3} $. The thick dashed line
  stands for the renormalized propagator of $ a^{\Phi} $,
  the thick solid line stands for the boson propagator.  The one loop diagram is not shown.}    

\vspace{0.25cm}

   For bosons, only one loop result is known:
\begin{equation}
   \Pi_{3} = N \frac{\pi}{8} (1+  g^{4} +  g^{8} + \cdots)
\end{equation}
    Note that both $ \Pi_{1} $ and $ \Pi_{3} $ are 
   even functions of $ \theta $.
  
  From Furry's theorem, one of the first non-vanishing diagram
  for $ \Pi_{2} $ is:

\vspace{0.25cm}

\epsfig{file=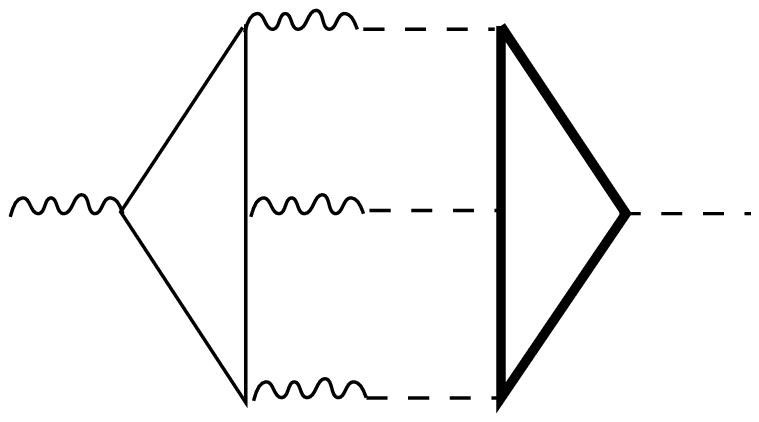,width=3.2in,height=0.8in,angle=0}

{\footnotesize {\bf Fig 5:} One of the four loop diagrams of $ \Pi_{2} $.
  the thin solid line stands for the fermion propagator,    
  the thick solid line for the boson propagator. All the other four loop diagrams can
  be obtained by shuffling the positions of the three bare propagator lines.}

\vspace{0.25cm}

  The series is $ \Pi_{2}= N( g^{6} + g^{10} + \cdots) $ which
  is an {\em odd} function of $ \theta $.

  From Eq.\ref{three}, $ \sigma^{\psi}, \sigma^{\Phi} $ are even functions
 of $ \theta $, but the mutual Hall drag conductivity is an {\em odd} function
 of $ \theta $. These are expected from P-H transformation.
 Under the P-H transformation of the
 vortex operator $ \Phi \rightarrow \Phi^{\dagger} $ in Eq.\ref{mutual},
  it can be shown
  that $ A^{\Phi}_{\mu}, \theta $ is equivalent to $ -A^{\Phi}_{\mu}, -\theta $.
  From Eq.\ref{matrix}, we reach the same conclusions. Specifically,
  $ \sigma^{H} $ takes opposite values for $ \theta=\pm 1/2 $,
  the periodicity under $ \theta \rightarrow \theta +1 $ is not preserved.
  Experimentally, the Hall drag conductivity can be detected by measuring
  the transverse voltage drop ( or transverse temperature drop for thermal
  conductivity ) of spinons due to the longitudinal driving of vortices.
  The Hall drag conductivity in double layer Quantum Hall systems has been
  investigated
  by several authors \cite{kun}. In double layer systems, the electrons
  in different layers are treated as two different species. There is a mutual
  CS interaction between the two species (both are fermions)
  which is directly responsible for
  this Hall drag conductivity, although the Coulomb interaction between the
  two species is responsible for the Coulomb ( longitudinal ) drag \cite{double}.
   This example shows that {\em no} external magnetic field is needed to produce
   a Hall effect ! However, $ \theta=\pm 1/2 $ are very special points,
   $ U(1) $ mutual
   CS term reduces to $ Z_{2} $ field, there should be {\em no } Hall drag
   conductivity just like $ \alpha= \pm 1/2 $ vortex leads to {\em no} Hall conductivity.

   Again, by using the large $ N $ result of \cite{chen,wen},
  we find the anomalous dimensions of the fermion and vortex to two loops:

\vspace{0.25cm}

\epsfig{file=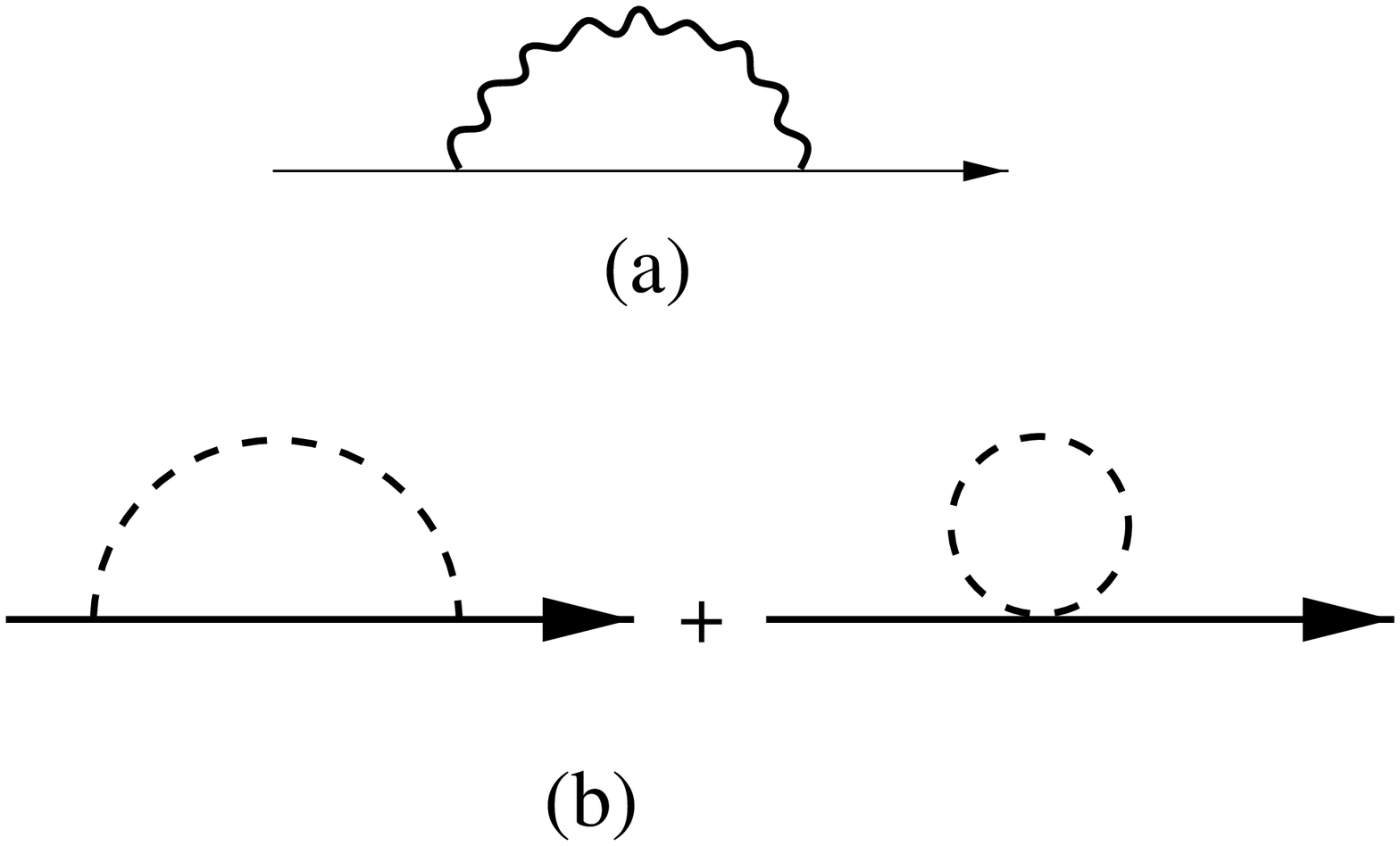,width=3.2in,height=0.9in,angle=0}

{\footnotesize {\bf Fig 6:} The two loop diagrams for the self energies of
  spinon (a) and vortex (b). The renormalized operators are used.}

\vspace{0.25cm}

     The results are:
\begin{eqnarray}
   \eta_{\psi} & = &  -\frac{g^{4}}{ 48 \pi^{2} }  \nonumber \\
   \eta_{\Phi} & = & \eta_{XY}  -\frac{g^{4}}{ 12 \pi^{2} } N 
\end{eqnarray} 
  where $ \eta_{XY} \sim 0.038 $ is the anomalous
  dimension for the 3d XY model\cite{recent}. 

  The correlation length exponent can also be calculated to two loops by 
  the insertion of the operator $ \Phi^{\dagger} \Phi $:
\vspace{0.25cm}

\epsfig{file=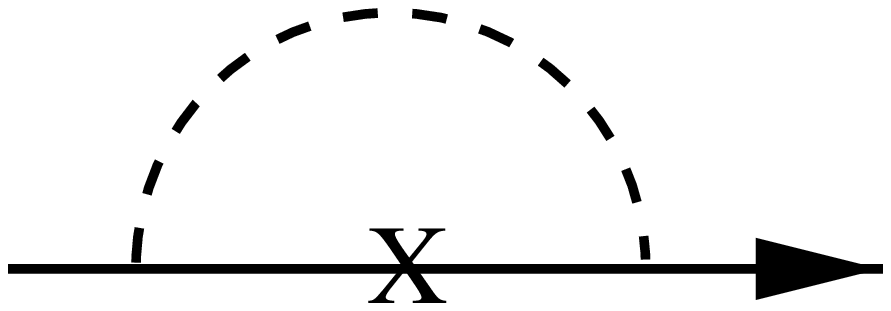,width=3.2in,height=0.8in,angle=0}

{\footnotesize {\bf Fig 7:} The cross stands for the operator insertion
  of $ \Phi^{\dagger} \Phi $.}

\vspace{0.25cm}

    The result is:
\begin{equation}
   \nu  =  \nu_{XY}  -\frac{g^{4}}{ 12 \pi^{2} } N
\end{equation}
  where $ \nu_{XY} \sim 0.672 $ is correlation length exponent for
  the 3d XY model\cite{recent}.

   It is instructive to go to dual representation of Eq.\ref{mutual}, namely
  go to the boson representation: 
\begin{eqnarray}
   {\cal L} &= & \psi^{\dagger}_{a} \gamma_{\mu} (\partial_{\mu}
    -i a^{\psi}_{\mu} -i A^{\psi}_{\mu} ) \psi_{a}   
   +  | ( \partial_{\mu} -i a^{\phi}_{\mu} ) \phi |^{2}
      + V( |\phi|)  \nonumber  \\
    &  + &  \frac{1}{4} ( f^{\phi}_{\mu \nu})^{2}+
    i a^{\Phi}_{\mu} 
    \epsilon_{\mu \nu \lambda } \partial_{\nu} ( a^{\phi}_{\lambda}
          -a^{\psi}_{\lambda}/\theta)
    -i A^{\Phi}_{\mu} \epsilon_{\mu \nu \lambda} \partial_{\nu}
      a^{\phi}_{\lambda}
\label{mutuald}
\end{eqnarray}
  where $ V(|\phi|)= m^{2}_{\phi} |\phi|^{2}+ g_{\phi} |\phi|^{4} +\cdots $,

    Integrating out $ a^{\Phi}_{\mu} $ leads to the constraint up to a pure
  gauge:
\begin{equation}
          a^{\psi}_{\mu}= \theta a^{\phi}_{\mu}
\label{cons}
\end{equation}

      Substituting the above constraint to Eq.\ref{mutuald} and setting
     $ a^{\phi}_{\mu} = a_{\mu} $, we find:
\begin{eqnarray}
   {\cal L} &= & \psi^{\dagger}_{a} \gamma_{\mu} (\partial_{\mu}
    -i \theta a_{\mu} -i A^{\psi}_{\mu} ) \psi_{a} 
     + \frac{1}{4}  f_{\mu \nu}^{2}     \nonumber  \\
  & + & | ( \partial_{\mu} -i a_{\mu} ) \phi |^{2} + V( |\phi|)
    -i A^{\Phi}_{\mu} \epsilon_{\mu \nu \lambda} \partial_{\nu}
      a_{\lambda}
\label{mutualb}
\end{eqnarray}

    The above Eq. indicates that fermions and bosons are coupled to the
  same gauge field whose dynamics is described by Maxwell term instead of
  C-S term. This Eq. is simply $ 2+1 $ dimensional combination of
   spinor QED and scalar QED. The RG analysis at $ 4-\epsilon $ by dimensional
  regularization is possible, because the marginal dimension of all the relevant
  couplings are 4. However a RG analysis directly at $ 2+1 $ dimension is formidable,
   the physical meaning and the exact marginality of $ \theta $ is 
  obscure in the boson representation. However they are evident in
  the dual vortex representation Eq.\ref{mutual}.  It is evident that
  there is no periodicity under $ \theta \rightarrow \theta +1 $ in
  Eq.\ref{mutualb}.

    If we perform duality  transformation again on Eq.\ref{mutualb} to
  go to the vortex representation, then
  we recover Eq.\ref{dual} upon neglecting the two Maxwell terms relative to
  the mutual C-S term.

    In the next subsection, setting the two source terms vanishing, we
   look at the properties of the different phases on the two sides of
   this quantum critical point.
\subsection{ Disorder and superconducting phases}
   In the disordered phase, the vortex condense $ < \Phi > = \Phi_{0}  $
   which generates a mass term for $ a^{\Phi}_{\mu} $ in Eq.\ref{mutual}
\begin{equation}
  \frac{\Phi^{2}_{0}}{2} ( a^{\Phi}_{\mu} )^{2}_{t}
\end{equation}
     where the subscript $ t $ means transverse projection.

    Integrating out the massive $ a^{\Phi}_{\mu} $ leads to a Maxwell term
  for $ a^{\psi}_{\mu} $:
\begin{equation}
   {\cal L} =  \psi^{\dagger}_{a} \gamma_{\mu} (\partial_{\mu}
    -i  a^{\psi}_{\mu} ) \psi_{a} + \frac{1}{4 \Phi^{2}_{0}
} ( f^{\psi}_{\mu \nu} )^{2}
\label{QED}
\end{equation}

    This is simply $ 2+1 $ dimensional spinor QED which was studied
  by large N expansion in \cite{qed}. In fact, we reach the same
  description from the boson representation Eq.\ref{mutualb}. Because
  in the disordered phase, the boson $ \phi $ is massive, therefore can be
  integrated out, it generates the Maxwell term $ \frac{1}{4 m_{\phi}}
     f^{2}_{\mu \nu} $ which dominates over the existing {\em non-critical}
   Maxwell term. We reach Eq.\ref{QED} after identifying $ m_{\phi} \sim
  \Phi^{2}_{0} $. 

     In the superconductor phase, the vortex $ \Phi $ is massive, therefore
  can be integrated out, it leads to the old Maxwell term 
    $  \frac{1}{4 m_{\Phi}} ( f^{\Phi}_{\mu \nu})^{2} $.
       Integrating out $ a^{\Phi} $ generates a mass term: 
\begin{equation}
   {\cal L} =  \psi^{\dagger}_{a} \gamma_{\mu} (\partial_{\mu}
    -i a^{\psi}_{\mu}) \psi_{a} 
     +\frac{m_{\Phi}}{2} (a^{\psi}_{\mu})_{t}^{2}
\label{free}
\end{equation}
    where $ m_{\Phi} $ is the mass of the vortex.

   The Dirac fermions become free.  In fact, we reach the
  same description from the boson representation Eq.\ref{mutualb}. Because
   in superconductor phase, the boson $ \phi $ condense, therefore generates
   a mass term $ \frac{\phi^{2}_{0}}{2} (a_{\mu})^{2}_{t} $ which renders
   the Maxwell term ineffective. We reach the same conclusion from both
   sides by identifying $ m_{\Phi} \sim \phi^{2}_{0} $.

    In short, in the disordered phase, the system is described by spinor
    QED Eq.\ref{QED}; in the superconductor side, by free Dirac fermion. 
    We can view the transition as the simplest
    confinement-deconfinement transition.
    In the confined ( disordered ) phase, the boson and fermion are confined 
    together by the fluctuating gauge field. In the deconfined ( superconductor)
    phase, the boson condensed, the fermion becomes free. There is a line
    of fixed point governed by the mutual statistical angle $ \theta $
    separating the two phases.
\section{ The effect of charge fluctuation}
  In this section, we try to investigate the effect of charge fluctuation on
  the fixed line characterized by the statistical angle $ \theta $ discussed
  in the last section.
  In the last section, we pointed out the periodicity of $ Z_{2} $ gauge
 CS term on the {\em lattice} is not preserved
 by $ U(1) $ gauge CS term in the {\em continuum}.
 In this section, we try to consider the combined effects of
  $ Z_{2} $ mutual statistical gauge fluctuation and $ U(1) $ electrical
  gauge fluctuation. After replacing the $ Z_{2} $ discrete gauge
  theory by a $ U(1) $ mutual CS theory, we treat $ U(1) $ mutual CS 
  theory and $ U(1) $ electrical gauge theory on the equal footing.
  Although the replacement leads to somewhat misleading conclusions,
  the heuristics derivation suggests that the condensation of
  $ hc/2e $ vortex condensation indeed leads to the confinement of
  spinon and chargon into Cooper pair and electron, in contrast to
  the condensation of $ hc/e $ vortex.  We do not
  intend to understand the critical behaviors of a combined $ Z_{2} $ and
  $ U(1) $ theory in this paper.
  We simply stress that $ Z_{2} $ periodicity should be treated correctly to
  understand the critical behaviors of a theory with both discrete $ Z_{2} $
  and continuous $ U(1) $ gauge fields. 
\subsection{ Quantum critical point}
  Putting $ a^{\Phi}_{\mu} \rightarrow 
      a^{\Phi}_{\mu} -a_{\mu} $ in Eq.\ref{dual}, we get
\begin{eqnarray}
   {\cal L} &= & \psi^{\dagger}_{a} \gamma_{\mu} (\partial_{\mu}
    -i a^{\psi}_{\mu}) \psi_{a}  +
   | ( \partial_{\mu} -i a^{\Phi}_{\mu} ) \Phi |^{2}
   + V( |\Phi|)   \nonumber  \\
    & + & \frac{i}{ 2 \pi \theta} a^{\psi}_{\mu} 
    \epsilon_{\mu \nu \lambda } \partial_{\nu} a^{\Phi}_{\lambda}
      - i a_{\mu} \epsilon_{\mu \nu \lambda } \partial_{\nu} (
     a^{\psi}_{\lambda}/\theta + A^{eff}_{\lambda} ) +
     \frac{1}{4} f^{2}_{\mu \nu}
\end{eqnarray}

    Integrating out the electric gauge field $ a_{\mu} $ leads to:
\begin{eqnarray}
   {\cal L} &= & \psi^{\dagger}_{a} \gamma_{\mu} (\partial_{\mu}
    -i a^{\psi}_{\mu}) \psi_{a}  +
   | ( \partial_{\mu} -i a^{\Phi}_{\mu} ) \Phi |^{2}
   + V( |\Phi|)   \nonumber  \\
    & + & \frac{i}{ 2 \pi \theta} a^{\psi}_{\mu} 
    \epsilon_{\mu \nu \lambda } \partial_{\nu} a^{\Phi}_{\lambda}
     +\frac{1}{2} (a^{\psi}_{\mu}/\theta + A^{eff}_{\mu} )_{t}^{2}
\label{vortexm} 
\end{eqnarray}

    Comparing to Eq.\ref{mutual},
   it is easy to see that the charge fluctuation leads to a mass term for
   the gauge field $ a^{\psi}_{\mu} $.
   Shifting $ a^{\psi}_{\mu}/\theta + A^{eff}_{\mu} \rightarrow a_{\mu} $ 
   and adding the gauge fixing term
    $ \frac{1}{2 \alpha} ( \partial_{\mu} a _{\mu} )^{2} $, we can integrate
    out the {\em massive } gauge field $ a_{\mu} $ in Lorenz gauge $
    \alpha=0 $ and find:
\begin{eqnarray}
   {\cal L} &= & \psi^{\dagger}_{a} \gamma_{\mu} (\partial_{\mu}
    -i \theta A^{eff}_{\mu}) \psi_{a}  +
   | ( \partial_{\mu} -i a^{\Phi}_{\mu} ) \Phi |^{2}
   + V( |\Phi|)   \nonumber  \\
    & + & \frac{1}{ 4} ( f^{\Phi}_{\mu \nu} )^{2} 
   -i ( A^{eff}_{\mu} + \theta j^{s}_{\mu} )
     \epsilon_{\mu \nu \lambda } \partial_{\nu} 
   a^{\Phi}_{\lambda} + (j^{s}_{\mu})^{2}_{t}
\end{eqnarray}

  Note the Maxwell term for $ a^{\Phi} $ is generated by the integration
  over the massive $ a_{\mu} $.

  Setting $ A_{\mu}=0 $, integrating out the fermions only leads to
 higher derivative terms than the Maxwell term:
\begin{equation}
   {\cal L} =  | ( \partial_{\mu} -i a^{\Phi}_{\mu} ) \Phi |^{2}
   + V( |\Phi|)   \nonumber  \\
     +  \frac{1}{ 4} ( f^{\Phi}_{\mu \nu} )^{2} + \cdots 
\end{equation}
  where $ \cdots $ means higher than second order derivatives.
 Therefore the vortex and fermion are asymptotically decoupled.
   It indicates that the charge fluctuation neglected in the last section
   destroy the fixed line characterized by $ \theta $. However,
   very different conclusions are reached in Ref.\cite{guess}.
   We believe that the authors failed to treat the charge gauge
   field fluctuation correctly.

  Just like the last section, it is instructive to go to dual
  representation of Eq.\ref{vortexm}, namely
  go to the boson representation: 
\begin{eqnarray}
   {\cal L} &= & \psi^{\dagger}_{a} \gamma_{\mu} (\partial_{\mu}
    -i a^{\psi}_{\mu} ) \psi_{a}   
   +  | ( \partial_{\mu} -i a^{\phi}_{\mu} ) \phi |^{2}
      + V( |\phi|)  \nonumber  \\
    &  + &  \frac{1}{4} ( f^{\phi}_{\mu \nu})^{2}+
    i a^{\Phi}_{\mu} 
    \epsilon_{\mu \nu \lambda } \partial_{\nu} ( a^{\phi}_{\lambda}
          -a^{\psi}_{\lambda}/\theta)  \nonumber  \\
     & +  &\frac{1}{2} (a^{\psi}_{\mu}/\theta + A^{eff}_{\mu} )_{t}^{2}
\label{charged}
\end{eqnarray}

   Integrating out $ a^{\Phi}_{\mu} $ leads to the {\em same}
  constraint as Eq.\ref{cons} up to a pure gauge:
\begin{equation}
          a^{\psi}_{\mu}= \theta a^{\phi}_{\mu}
\end{equation}

      Substituting the above constraint to Eq.\ref{charged} and setting
     $ a^{\phi}_{\mu} = a_{\mu} $, we find:
\begin{eqnarray}
   {\cal L} &= & \psi^{\dagger}_{a} \gamma_{\mu} (\partial_{\mu}
    -i \theta a_{\mu} ) \psi_{a}   
   +  | ( \partial_{\mu} -i a_{\mu} ) \phi |^{2}
      + V( |\phi|)  \nonumber  \\
   & + & \frac{1}{4}  f_{\mu \nu}^{2}
     +\frac{1}{2} (a_{\mu} + A^{eff}_{\mu} )_{t}^{2}
\label{chargeb}
\end{eqnarray}

   Comparing to Eq.\ref{mutualb}, the only difference is that the gauge field
  acquires a mass due to the charge fluctuation which renders
  the Maxwell term ineffective. Up to irrelevant couplings,
  we can safely set $ a_{\mu}= A^{eff}_{\mu} $ in
  the above equation and find
\begin{equation}
   {\cal L} =  \psi^{\dagger}_{a} \gamma_{\mu} (\partial_{\mu}
    -i 2 \theta A_{\mu} ) \psi_{a}   
   +  | ( \partial_{\mu} -i 2 A_{\mu} ) \phi |^{2}
      + V( |\phi|) + \cdots 
\label{decouple}
\end{equation}
  where $ \cdots $ means the irrelevant couplings between bosons and fermions.

  The above Eq. leads to in-consistent conclusion that the original
  charge neutral spinon $\psi $
  becomes charged: carrying charge $ e $ for $ \theta=1/2 $ and $ -e $
  for $ \theta=-1/2 $. Therefore we also reach ambiguous conclusions for $ \theta
  =1/2 $ and $ \theta=-1/2 $. The reason is that we replace the $ Z_{2} $ CS
  gauge theory on the lattice which explicitly respects the equivalence
  between $ \theta=1/2 $ and $ \theta=-1/2$ by the $ U(1) $ CS theory
  which does not respect this equivalence in the continuum limit. Identifying
  $ a^{\psi}_{\mu} $ which is a discrete $ Z_{2} $ gauge field with
  $ A_{\mu} $ which is a $ U(1) $ gauge field is a direct consequence of
  this problematic replacement.
\subsection{ Disordered and superconducting phases}
  We follow the discussions in the previous section. In the disordered phase,
  the vortex condense $ < \Phi > = \Phi_{0}  $
  which generates a mass term for $ a^{\Phi}_{\mu} $ in Eq.\ref{vortexm}.
  Integrating out the massive $ a^{\Phi}_{\mu} $ leads to a Maxwell term
  for $ a^{\psi}_{\mu} $.
\begin{equation}
   {\cal L} =  \psi^{\dagger}_{a} \gamma_{\mu} (\partial_{\mu}
    -i a^{\psi}_{\mu}) \psi_{a} 
     +\frac{1}{4 \Phi^{2}_{0}} (f^{\psi}_{\mu \nu})^{2}
     +\frac{1}{2} (a^{\psi}_{\mu}/\theta + A^{eff}_{\mu} )_{t}^{2}
\end{equation}

   The Dirac fermions become free and carry spin $ 1/2 $ and charge
   $ 2 \theta e $.
   In fact, we reach the same conclusion from the boson representation
   Eq.\ref{chargeb}.

   In the superconductor phase, the vortex $ \Phi $ is massive, therefore
   can be integrated out, it leads to the old Maxwell term 
   $  \frac{1}{4 m_{\Phi}} ( f^{\Phi}_{\mu \nu})^{2} $.
   Integrating out $ a^{\Phi} $ generates a mass term for
   $ a^{\psi}_{\mu} $:
\begin{equation}
   {\cal L} =  \psi^{\dagger}_{a} \gamma_{\mu} (\partial_{\mu}
    -i a^{\psi}_{\mu}) \psi_{a} 
     +\frac{m_{\Phi}}{2} (a^{\psi}_{\mu})_{t}^{2}
     +\frac{1}{2} (a^{\psi}_{\mu}/\theta + A^{eff}_{\mu} )_{t}^{2}
\label{con}
\end{equation}
    where $ m_{\Phi} $ is the mass of the vortex.

   Diagonizing the last two mass terms leads to a {\em continuously changing
  charge }.
  In fact, we reach the same description from the boson representation
   Eq.\ref{chargeb}. Because
   in superconductor phase, the boson condense $ <\phi> =\phi_{0} $,
   therefore generates a mass term for $ a_{\mu} $:
\begin{equation}
   {\cal L} =  \psi^{\dagger}_{a} \gamma_{\mu} (\partial_{\mu}
    -i \theta a_{\mu} ) \psi_{a}   
    +  \frac{\phi^{2}_{0}}{2} (a_{\mu} )_{t}^{2}
     +\frac{1}{2} (a_{\mu} + A^{eff}_{\mu} )_{t}^{2}
\end{equation}
     Which is essentially the same as Eq.\ref{con} after the identification
   $ m_{\Phi} \sim \phi^{2}_{0} $.

   Again, the pathological claims are due to the artifact of replacing the
   $ Z_{2} $ mutual
   statistical gauge theory by a $ U(1) $ mutual CS theory, then treating
   $ Z_{2} $ gauge field and $ U(1) $ electrical gauge field on the same
   footing. Although all these results are the artifacts of this replacement,
   they do suggest that the mutual statistical interaction between $ hc/2e $
   vortex and spinons leads to the confinement of spinon and chargon into
   electron and Cooper pair.

\section{discussions and conclusions}

  The original singular gauge transformation \cite{and,sing} was proposed
  for static vortices. In this paper, we extend the singular gauge
  transformation to moving
  vortices generated by quantum fluctuation. By making a close
  analogy to the conventional singular gauge transformation of FQH system, we
  perform  singular gauge transformations
  attaching flux of moving vortices to quasi-particles or vice versa. 
  Just like conventional singular gauge transformation leads to
  conventional CS term, the two mutual singular gauge transformations
  lead to mutual CS term. In this way, we propose a intuitive and
  physical transparent approach to bring out explicitly
  the underlying physics associated with the condensation of the
  $ hc/2e $ vortices.

 Based on the earlier work by Balents, Fisher and Nayak \cite{balents},
 Senthil and Fisher developed $ Z_{2} $
 gauge theory \cite{z2,guess} to study quasi-particles
 coupled to vortices generated by quantum fluctuations.
 By breaking electrons and Cooper pairs into smaller constitutes: chargons with
 charge $ e $, spin $ 0 $ and spinons with charge $ 0 $,spin $ 1/2 $,
 SF introduced a local $ Z_{2} $ gauge degree of
 freedom to constrain the
 Hilbert space to the original one. The effective
 action describes both chargons and spinons coupled to local fluctuating
 $ Z_{2} $ gauge theory with a doping dependent Berry phase term.
 By the combination of standard duality transformation of 3 dimensional
 XY model and that of $ Z_{2} $ gauge theory, the action is mapped into a
 dual vortex representation where the $ hc/2e $ vortices and spinons 
 are coupled by a mutual $ Z_{2} $ CS gauge theory. As usual
 vortices in XY model, the $ hc/2e $ vortices also couple to a fluctuating
 $ U(1) $ gauge field which mediates the long-range logarithmic interaction
 between the vortices. Starting from the dual representation,
 the authors in Ref.\cite{guess}
 studied a transition from d-wave superconductor to confined Mott
 insulator driven by
 the condensing of $ hc/2e $ vortices. In order to study the critical
 behaviors of this particular confinement and de-confinement transition,
 they replaced the $ Z_{2} $ mutual CS theory on the lattice by $ U(1) $
 mutual CS theory in the continuum and performed renormalization Group (RG) 
 analysis. Unfortunately, some of their RG analysis are incorrect as
 demonstrated in Sec.IV.

  It was shown in the text that our resulting general effective action
  Eq.\ref{dual} at the two special statistical
  angles $ \theta=\pm 1/2 $ is essentially equivalent to
  the dual vortex representation of $ Z_{2} $ gauge theory developed
  by SF, except it
  also bring out explicitly the Volovik effect which was  {\em missed} in
  SF's $ Z_{2} $ gauge theory.

  As explicitly stressed in this paper, the $ U(1) $
  mutual CS theory in the continuum does not have the required periodicity under
  $ \theta \rightarrow \theta +1 $. In order to preserve this periodicity
  when the mutual statistical angle takes special values
  $ \theta = \pm 1/2 $, a $ Z_{2} $
  mutual CS theory must be enforced on the lattice.
  Again, two ways of regularization may lead to different conclusions.
  In fact, the periodicity of the conventional CS theory under
  $ \theta \rightarrow \theta+2 $ is also a very intricate issue.
  On the one hand,
  the perturbative RG expansion in terms of the statistical angle $ \theta $ in
  the continuum limit in Refs.\cite{wen,boson,subir} does not
  have this periodicity.
  On the other hand, properly regularized on the lattice,
  the CS theory does have
  this periodicity \cite{kiv}. 
  The two different regularization do lead to different conclusions on the
  Quantum Hall transitions. It also leads to notorious Hall conductivity
  difficulty at $ \nu=1/2 $ \cite{not}.
  Although we are unable to solve the critical behaviors
  of this combined $ Z_{2} $ and $ U(1) $ gauge fields coupled to spinons
  and $ hc/2e $ vortices, we stressed the importance of putting the theory on
  the lattices to keep the periodicity of $ Z_{2} $ mutual CS term and
  shed considerable lights on the physical picture and structure 
  of the quantum transitions driven by $ hc/2e $ vortices.
  We suggest that it may be possible to trace out the $ Z_{2} $
  gauge fluctuation on the lattice first to get an effective action
  without the discrete gauge fields, then take continuum limit of it
  and perform RG calculation.  Further work is needed to
  sort out all the possible phases and to understand the critical
  behaviors of the phase transitions between these phases.
  
 This work was supported by NSF Grant No. PHY99-07949,
 university of Houston and DMR-97-07701.
 We thank M. P. A. Fisher, S. Girvin, S. Kivelson,
 P. A. Lee, S, Sachdev, T. Senthil, Z. Tesanovic, Y. S. Wu and
 Kun Yang for helpful discussions.

\appendix

\section{ The most general form of the gauge propagators}

    Adding the gauge fixing terms $ \frac{1}{2 \alpha}
   ( (\partial_{\mu} a^{\psi}_{\mu} )^{2}+
    (\partial_{\mu} a^{\Phi}_{\mu} )^{2}) $ to Eq.\ref{mutual}, we
   can find the gauge field propagators by inverting the matrix 
\begin{equation}
             \left( \begin{array}{cc}
    \Pi_{1}(k) k ( \delta_{\mu \nu} - \frac{ k_{\mu} k_{\nu} }{k^{2}} )
     + \frac{ k_{\mu} k_{\nu} }{\alpha} &
    \Pi_{2}(k) \epsilon_{\mu \nu \lambda} k_{\lambda}   \\
    \Pi_{2}(k) \epsilon_{\mu \nu \lambda} k_{\lambda}   &
    \Pi_{3}(k) k ( \delta_{\mu \nu} - \frac{ k_{\mu} k_{\nu} }{k^{2}} )
     + \frac{ k_{\mu} k_{\nu} }{\alpha} \\
			\end{array}   \right )
\end{equation}
    
     The results are:
\begin{eqnarray}
    (G_{\psi \psi})_{\mu \nu} & = & \frac{ \alpha k_{\mu} k_{\nu} }{ k^{4} }
    + \frac{\Pi_{3}}{ \Pi_{1} \Pi_{3} + \Pi^{2}_{2} } \frac{1}{k}
    ( \delta_{\mu \nu} - \frac{ k_{\mu} k_{\nu} }{k^{2}} ) 
               \nonumber  \\
    (G_{\psi \Phi})_{\mu \nu} & = & (G_{\Phi \psi})_{\mu \nu} = 
     \frac{\Pi_{2}}{ \Pi_{1} \Pi_{3} + \Pi^{2}_{2} } (- \frac{
     \epsilon_{\mu \nu \lambda} k_{\lambda} }{k^{2}} )
                    \nonumber  \\
    (G_{\Phi \Phi})_{\mu \nu} & = & \frac{ \alpha k_{\mu} k_{\nu} }{ k^{4} }
    + \frac{\Pi_{1}}{ \Pi_{1} \Pi_{3} + \Pi^{2}_{2} } \frac{1}{k}
    ( \delta_{\mu \nu} - \frac{ k_{\mu} k_{\nu} }{k^{2}} )  
 \end{eqnarray}

    In the Landau gauge $ \alpha=0 $ and putting $ \vec{k}=0 $, we recover
   the results calculated in Sec. III.

\section{ Application to static disordered vortex array}

  In Ref.\cite{static}, Ye pointed out that the long range Logarithmic
  interaction between vortices suppress the superfluid velocity fluctuation,
  but does not affect the internal gauge field fluctuation. He concluded
  that the quasi-particle scattering from the random gauge field dominate
  over that from the superfluid velocity ( the Volovik effect). In this
  appendix, by using the formalism developed in the main text, we provide
  additional evidence for this conclusion.

    For random pinned vortices, the static vortices do not feel
   the statistical gauge field $ a^{\Phi}_{\mu} $ in Eq.\ref{dual}.
   This corresponds to a mass term for this gauge field. Adding the mass term
  to the equation and integrating out $ a^{\Phi}_{\mu} $, it generates
  a Maxwell term for $ a^{\psi}_{\mu} $ \cite{random}:
\begin{eqnarray}
   {\cal L} &= & \psi^{\dagger}_{a} \gamma_{\mu} (\partial_{\mu}
    -i a^{\psi}_{\mu} ) \psi_{a}  
   +  \frac{1}{ 4 } ( f^{\psi}_{ \mu \nu})^{2}   \nonumber  \\ 
  & + & | ( \partial_{\mu} -i a_{\mu} ) \Phi |^{2}
      + V( |\Phi|) +  \frac{1}{ 4 } f^{2}_{ \mu \nu} 
   -i A^{eff}_{\mu} \epsilon_{\mu \nu \lambda } \partial_{\nu} a_{\lambda}
\end{eqnarray}

    In the above Eq. the averages over $ a^{\psi}_{\mu}, a_{\mu}, \Phi $
  should be understood as quenched instead of annealed averages.
    As discussed in the main text, the long-range logarithmic interaction
 between vortices represented by the Maxwell term suppress the fluctuation of
 the superfluid velocity. The quasi-particle is moving in a long-range
 correlated random magnetic field and short-range correlated scaler potential,
 in consistent with the physical picture in Ref.\cite{static}.

\section{ $ hc/e $ vortices, Spin-charge separation and Mott insulator }

 In sec III, neglecting
  the charge fluctuation, we concentrated on the effect of mutual C-S term.
  In this appendix, we do the opposite, namely neglecting the mutual C-S term,
  but concentrate on the effect of charge fluctuation.
 This is realized by using neutral-like transformation, namely
  setting $ \phi_{A}= \phi_{B}= \phi/2 $ which is single-valued in the 
  resulting $ hc/e $ vortices.
 The double strength vortex and its stability was investigated in
 Ref.\cite{own}, the associated spin-charge separation
  has been discussed extensively in Refs.\cite{balents,z2}. In this appendix,
  to contrast with $ hc/2e $ vortices discussed in the main text,
  we review their results, fill in some details and also get the new result
  on the expression of current operator $ I_{\mu} $ in the superconducting phase. 

     Putting $ \phi_{A}=\phi_{B}=\phi/2 $, namely setting $ a_{\mu}=0 $
    in Eq.\ref{act} leads to: 
\begin{eqnarray}
   {\cal L} &= & \psi^{\dagger}_{1a} [ \partial_{\tau} 
    + v_{f} p_{x} \tau^{3} + v_{2} p_{y} \tau^{1} ]
    \psi_{1a} +  (1 \rightarrow 2, x \rightarrow y )
    \nonumber   \\
     & + & \frac{K}{2} ( \partial_{\mu} \phi - A^{eff}_{\mu} )^{2}
\label{acta}
\end{eqnarray}
    where the effective gauge field is $ A^{eff}_{\mu}= A_{\mu}-
    K^{-1}J_{\mu} $. There is no internal gauge
    field $ a_{\mu} $,
   because double vortices do not scatter the quasi-particles.

    Performing duality transformation to vortex representation leads to
\begin{eqnarray}
   {\cal L} &= & \psi^{\dagger}_{a} \gamma_{\mu} \partial_{\mu} \psi_{a}  +
   | ( \partial_{\mu} -i a_{\mu} ) \Phi |^{2}
   + V( |\Phi|)   \nonumber  \\
    & + & \frac{1}{ 4}  f_{\mu \nu}^{2} 
   -i  A^{eff}_{\mu} \epsilon_{\mu \nu \lambda } \partial_{\nu} 
   a_{\lambda} -\mu \epsilon_{ij} \partial_{i} a_{j} 
\label{double}
\end{eqnarray}
   where the last ( Berry phase ) term can be absorbed into  $ A^{eff}_{\mu} $
  by redefining $ A^{eff}_{\mu} \rightarrow A^{eff}_{\mu} +
   i \mu \delta_{\mu 0} $.
    Comparing with Eq.\ref{dual}, there is only charge fluctuation.
 There is no mutual C-S term, namely $\theta=0 $ or any {\em integer }. 

    The duality transformation to boson representation is
\begin{eqnarray}
   {\cal L} &= & \psi^{\dagger}_{a} \gamma_{\mu} \partial_{\mu} \psi_{a}   
   +  | ( \partial_{\mu} -i a^{\phi}_{\mu} ) \phi |^{2}
      + V( |\phi|)  \nonumber  \\
    &  + &  \frac{1}{4} ( f^{\phi}_{\mu \nu})^{2}+
    i a_{\mu} 
    \epsilon_{\mu \nu \lambda } \partial_{\nu} ( a^{\phi}_{\lambda}
          -A^{eff}_{\lambda}/\theta)
    + \frac{1}{4} f_{\mu \nu}^{2}
\end{eqnarray}

    Integrating out the electric gauge field leads to
\begin{eqnarray}
   {\cal L} &= & \psi^{\dagger}_{a} \gamma_{\mu} \partial_{\mu} \psi_{a}   
   +  | ( \partial_{\mu} -i a^{\phi}_{\mu} ) \phi |^{2}
      + V( |\phi|)      \nonumber  \\
    &  + &  \frac{1}{4} ( f^{\phi}_{\mu \nu})^{2}
    + \frac{1}{2} ( a^{\phi}_{\mu} -A^{eff}_{\mu})^{2}_{t}
\end{eqnarray}

    Setting  $ a^{\phi}_{\mu}= A^{eff}_{\mu} $ up to irrelevant term leads to
\begin{equation}
   {\cal L} = \psi^{\dagger}_{a} \gamma_{\mu} \partial_{\mu} \psi_{a}   
   + n_{0} \phi^{\dagger} \partial_{\tau} \phi
   +  | ( \partial_{\mu} -i  A_{\mu} ) \phi |^{2}
      + V( |\phi|) + \cdots 
\end{equation}
  where the second term is the  Berry phase term and
 $ \cdots $ means the irrelevant couplings between bosons and fermions.
  The above Eq. is essentially the same as Eq.\ref{acta}.
  Bosons and fermions are asymptotically decoupled. The
   fermions (spinons) carry  only spin 1/2, the bosons ( holons)
   carry only charge $ e $.  There is spin-charge separation.
   The long-range Coulomb interaction could be incorporated by adding the dynamic term
   for the time component of the gauge field $ \frac{1}{2} k | A_{0}(k) |^{2} $ \cite{coul}.
   Without the fermionic part,
   the action is the same as the superconductor-insulator transitions in dirty metals
   studied in \cite{sit}, except here condensed is a charge $ e $ boson instead of a Cooper pair.

   In the underdoped regime, the vortex condenses $ < \Phi > = \Phi_{0}  $
   which generates a mass term for $ a_{\mu} $ in Eq.\ref{double}
   $ \frac{1}{2} \Phi^{2}_{0} ( a_{\mu} )^{2}_{t} $.
    Integrating out the massive $ a_{\mu} $ leads to a Maxwell term
  for $ A^{eff}_{\mu} =A_{\mu}-K^{-1} J_{\mu} $ from which we can identify
  the automatically conserved electric current 
   $ I_{\mu} = \frac{1}{ \Phi^{2}_{0}}
   ( \partial^{2}_{\nu} J_{\mu}
   - \partial_{\mu} \partial_{\nu} J_{\nu} ) $ \cite{balents}.

    In the superconductor phase, the vortex $ \Phi $ is massive, therefore
    can be integrated out, it leads to the old Maxwell term 
    $  \frac{1}{4 m_{\Phi}} ( f_{\mu \nu})^{2} $.
    Integrating out $ a_{\mu} $ generates a mass term  
   $ \frac{1}{2} m_{\Phi}( A^{eff}_{\mu})^{2}_{t} $. From the mass term,
   we can identify the automatically conserved electric current 
   $ I_{\mu}= m_{\Phi} ( J_{\mu} )_{t} $. Note that although the quasi-particle
    electric current is not conserved itself, the total electric current
    $ I_{\mu}= m_{\Phi} ( \delta_{\mu \nu}-\frac{q_{\mu} q_{\nu} }{q^{2}} )
         J_{\nu}(q) $ is conserved.

\end{multicols}

\begin{references}
\bibitem{curve} In this paper, we neglect the curvature term which
  may be important for the thermal Hall conductance in the presence
  of external magnetic field. For details, see Ref.\cite{static}.
\bibitem{break} We can also directly add Eq.\ref{linear} and
  Eq.\ref{tilde} without
  performing this singular gauge transformation and then take continuum limit.
  But the resulting continuum action breaks spin $ SU(2) $ symmetry.
\bibitem{hole} If the quasi-particle is hole-like, then $ a_{\mu} 
   \rightarrow -a_{\mu} $, but $ v_{\mu} $ which is gauge invariant
   stays the same. We reach the same conclusion.
\bibitem{exchange} Note that under this transformation, the Time (T) reversal
   symmetry of $ c $ becomes Particle and Hole (PH) symmetry of $ d $ and
   vice versa.  In the following, we use the language in the
   $ d $ representation.
\end{references}
\end{document}